\documentclass[preprint,preprintnumbers,showpacs,prb,amsmath,amssymb,floatfix,showpacs]{revtex4}

\usepackage{graphicx}
\usepackage{dcolumn}
\usepackage{bm}
\begin{document}
\preprint{\today}
\title{Phase behaviour of attractive and repulsive ramp fluids:
  integral equation and computer simulation studies}
\author{E.~Lomba, N.G.~Almarza, C.~Mart\'{\i}n and C.~McBride}
\affiliation{Instituto de Qu\'\i mica F\'\i sica Rocasolano, CSIC,  Serrano 119,
  E-28006, Madrid, Spain
}

\date{\today}
\begin{abstract}
Using computer simulations and a thermodynamically self consistent
 integral equation we investigate the phase behaviour 
 and thermodynamic anomalies of a fluid composed of spherical
 particles interacting via a two-scale ramp potential (a hard core
 plus a repulsive and an attractive ramp) and the corresponding purely repulsive
 model. Both simulation and integral equation results predict a
 liquid-liquid de-mixing  when attractive forces are present, in addition to
 a gas-liquid transition. Furthermore, a fluid-solid transition
 emerges in the neighbourhood of the liquid-liquid transition region, leading to a phase
 diagram with a somewhat complicated topology. This solidification at
 moderate densities is also present in the repulsive ramp fluid, thus 
 preventing fluid-fluid separation. 
\end{abstract}


\maketitle
\protect\section{Introduction}
The existence of a liquid-liquid (LL) equilibrium and density, diffusivity
and structural anomalies in single component fluids has attracted 
considerable interest in the last decade. The density and
diffusion anomalies present in  liquid water (i.e. the existence of a maximum in the density within the liquid phase, 
and an increase in diffusion upon compression, up to a certain point) 
have long since been known, and have been
reproduced by several of the existing interaction models\cite{JCP_1996_105_5099,PRL_2002_88_195701,PNAS_2005_102_16558,JCP_2005_123_044515,JCP_2005_123_144504}. Experimental evidence for
LL coexistence has been found for phosphorous\cite{NAT_2000_403_170}, triphenyl
phosphite\cite{PRL_2004_92_025701}, and
$n$-butanol\cite{JPCM_2005_17_L293}, and it has been suggested that
this LL equilibrium might be the
source of the anomalies encountered in water. Computer simulations  predict the
existence of LL equilibria not only in water\cite{JCP_1996_105_5099,PRL_2002_88_195701,JCP_2005_123_044515} but
also in other loosely coordinated fluids, such as
silicon\cite{NM_2003_2_739}, carbon\cite{PRB_2004_69_100101} and
silica\cite{PRE_2000_63_011202}. Whether these transitions physically exist or not is still open to
debate, since in most cases they correspond to supercooled states
which are rendered experimentally inaccessible by
crystallisation. However, closely related first order transitions
between low and high density amorphous phases have indeed been found
for water\cite{JCP_1993_100_5910}, silica\cite{PRL_2001_87_195501} and
germanium oxide\cite{JCP_1995_102_6851}. 

While the use of realistic models can
 provide reasonable explanations for the experimental
behaviour of a physical system, a more thorough description of the  mechanisms
underlying both LL phase transitions and density,
diffusivity and related anomalies can be acquired via the
study of simplified models. In the case of LL equilibrium and
polyamorphism in molecular fluids\cite{JPC_1996_100_8518} the simple
model of Roberts and Debenedetti\cite{JCP_1996_105_658} has 
successfully accounted for the behaviour of network forming
fluids\cite{PRL_1996_77_4386}. 
On the other hand, since the pioneering
work of Hemmer and Stell\cite{PRL_1970_24_1284,JCP_1972_56_4274} it is 
known that a simple spherically symmetric potential in which the
repulsive interaction has been softened (in this case by the addition
of a repulsive ramp to the hard core) can lead to the existence of
a second (LL)  critical point,  as long as  a first (liquid-vapour, LV) critical
point existed due to the presence of a long range attractive component in
the interaction potential.
As well as the ramp potential, other simple models
with two distinct ranges  of interaction, such as 
the hard-sphere square shoulder-square well potential studied by
Skibinsky et al.\cite{PRE_2004_69_061206}, have also been shown to exhibit
LL equilibria. 
Indeed the presence of two interaction ranges 
explains the competition between two locally preferred
structures (LPS) -- a LPS being defined as an arrangement of
particles which, for a given state point, minimises some local Helmholtz
energy \cite{JPCM_2003_15_S1077}. This competition between two LPS 
helps to rationalise the existence of  polyamorphism  and LL equilibria
in single component glassy systems and
fluids\cite{JPCM_2003_15_S1077}. 

More recently, the original model proposed by Hemmer and Stell has regained attention,
especially since Jagla\cite{JCP_1999_111_8980} stressed
the similarities between the behaviour of the Hemmer-Stell ramp
potential and the anomalous properties of liquid water.
This has been further explored by Xu et
al.\cite{PNAS_2005_102_16558,JPCM_2006_18_S2239} who analysed the
relationship between the LL transition and changes in the dynamic
behaviour of fluids interacting via a soft core ramp potential with
attractive dispersive interactions added. Gibson and
Wilding\cite{PRE_2006_73_061507} have recently presented an exhaustive
study of a series of ramp potentials exhibiting LL transitions and
density anomalies, whose relative position and stability with respect
to freezing might be tuned by judicious changes in the interaction
parameter. Moreover, Caballero and Puertas\cite{PRE_2006_74_051506}
have also focused on the relation 
between the density anomaly and the LL transition for this model
system by means of a Monte Carlo based perturbation approach. The aforementioned authors find
that, in this case, the density anomaly is absent when the range of
the attractive interaction is sufficiently small. 

In this work  we shall refer to this
interaction potential as the `attractive two-scale ramp potential'
(A2SRP). Additionally, when one considers the system without any
attractive contribution, i.e. a hard-sphere core plus a repulsive
ramp, as the `two-scale ramp potential' (2SRP). Although it has been found that the
system exhibits density anomalies, it is a possibility that LL transition is
preempted by crystallisation\cite{PRE_2005_72_021501}. In this purely
repulsive model, the relation between static and dynamic anomalies
has been explored in
detail\cite{PRE_2005_72_021501,JCP_2006_125_204501,JCP_2006_125_244502} (similar results
have been found for a dumbbell fluid with repulsive ramp site-site
interactions\cite{PRE_2006_73_061504}) and the connection between these
anomalies and structural order has been investigated by Yan et
al.\cite{PRL_2005_95_130604} with the aid of the Errington-Debenedetti
order map\cite{NAT_2001_409_318,JCP_2003_118_2256}. These authors
found that there is a region of structural anomalies (in which both
translational and orientational order decrease as density is
increased) that encapsulates the diffusivity and density anomaly region.
The intimate relationship between transport coefficients,
and the excess entropy
was made clear first by Rosenfeld \cite{PRA_1977_15_002545,JPCM_1999_11_05415},
and again later, specifically for the atomic diffusion,  by Dzugutov \cite{N_1996_381_00137_photocopy}.
Needless to say, a corollary of this relation is that any 
anomalous diffusion will be accompanied by an anomalous excess entropy.
Recently this link has been shown to hold
for both liquid silica and for the 2SRP model by Sharma et al. \cite{JCP_2006_125_204501},
and for the discontinuous core-softened model by Errington et al.\cite{JCP_2006_125_244502}. 

 The principal objective of this work is to extend our knowledge of the phase
behaviour of systems interacting via either attractive or purely repulsive two scale ramp
potentials (A2SRP and 2SRP). To that purpose 
exhaustive Monte Carlo calculations have been performed in order to determine the phase
boundaries of the gas, liquid and solid phases for both model systems up to
moderate densities --slightly beyond the high density branch of the  LL
equilibrium.  Self-consistent integral equation
calculations performed on the A2SRP model complement the Monte Carlo results
and agree qualitatively as to the location of the LL equilibria and
quantitatively for the LV equilibria. The location of Widom's line
(the locus of  heat capacity maxima) is obtained for both models. The
temperature of maximum density (TMD) curve is obtained for the
repulsive 2SRP model and correlated with the location of Widom's
line. The rich variety of phases present in these simple models will
be illustrated in the calculated phase diagrams. 

The structure of the paper is as follows; the model and
computational procedures used herein are introduced in Section II. In
Section III the most significant results are presented and discussed.
Finally, the main conclusions and future
prospects can be found in Section IV.

\section{Model and computational methods}
\begin{figure}
\includegraphics[width=8cm,clip]{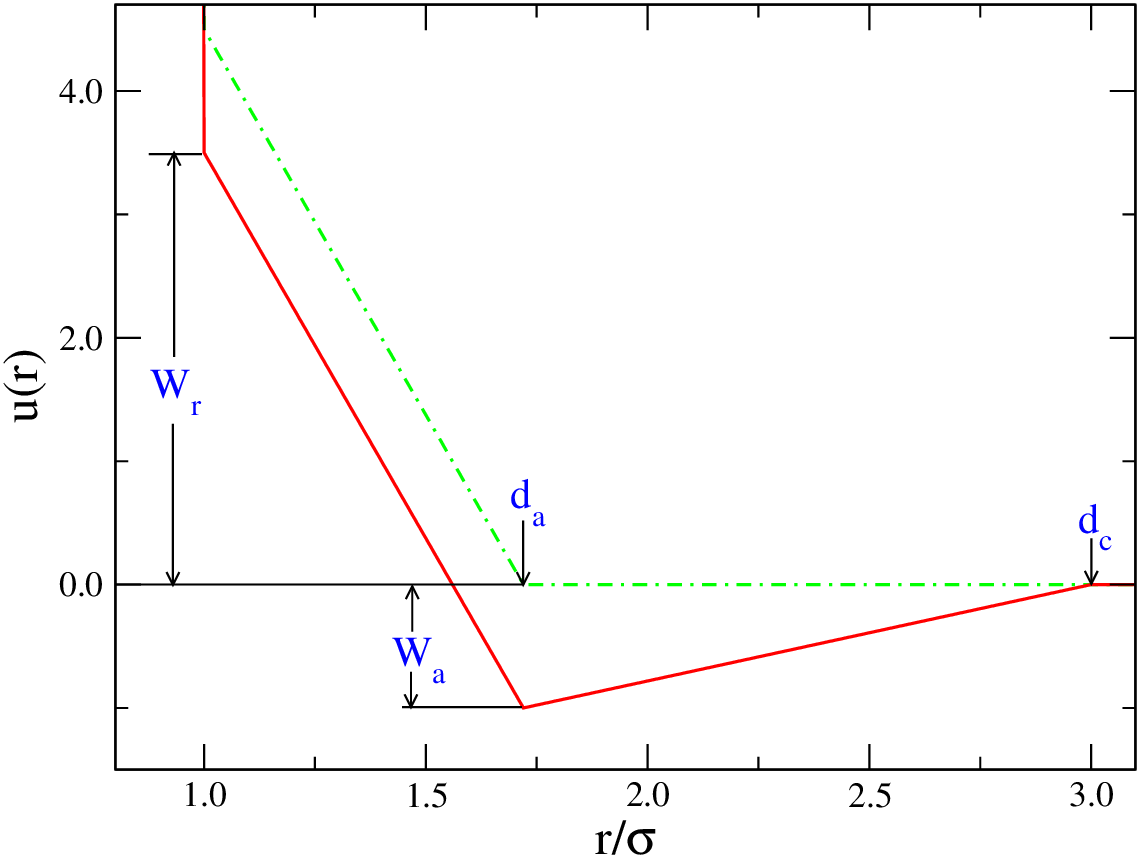}
\caption{Attractive two scale ramp (A2SRP, solid line)
  and repulsive two scale ramp (2SRP, dash-dotted line) potential
  models.}
\label{fig1}
\end{figure}

The first model system consists of hard spheres of diameter $\sigma$, with a
repulsive soft core and an attractive region. The interaction
potential reads 
\begin{equation}
u(r) = \left\{
\begin{array}{ll}
\infty & \mbox{if}\; r < \sigma \\
W_r - (W_r-W_a)\frac{r-\sigma}{d_a-\sigma} & \mbox{if}\;\sigma\leq r \leq
d_a \\
W_a - W_a\frac{r-d_a}{d_c-d_a} &  \mbox{if}\;d_a < r \leq d_c\\
0 &  \mbox{if}\; r > d_c
\end{array}
\right.
\label{pot}
\end{equation}
where $W_r > 0$ and $W_a < 0$.
The values used here for the parameterisation of the model are 
the same as those used in 
Ref. \onlinecite{PNAS_2005_102_16558}. Thus, energy units are
  defined by $|W_a|$, with the reduced temperature 
  $T^*=k_BT/|W_a|$ ($k_B$ being the Boltzmann constant) and we have set
  $W_r/|W_a|=3.5$. The units of length 
  are reduced with respect to the hard core  diameter, and so one has
  $d_a/\sigma=1.72$, $d_c/\sigma=3.0$ and the reduced density is
  $\rho^*=\rho\sigma^3$ as usual. This set of parameter values and
  Eq.(\ref{pot}) define the A2SRP model. For the purely repulsive
  system, we have chosen the corresponding repulsive potential that
  would result from a Weeks-Chandler-Anderson\cite{JCP_1971_54_5237}
  decomposition of  Eq.(\ref{pot}), namely
\begin{equation}
u_r(r) = \left\{
\begin{array}{ll}
\infty & \mbox{if}\; r < \sigma \\
W_r -W_a - (W_r-W_a)\frac{r-\sigma}{d_a-\sigma} & \mbox{if}\;\sigma\leq r \leq
d_a \\
0 &  \mbox{if}\;r > d_a
\end{array}
\right.
\label{potr}
\end{equation}
In Figure \ref{fig1} both the attractive and purely repulsive
interactions are illustrated. 

\subsection{The self consistent integral equation approach}
The  integral equation calculations are based on the Ornstein-Zernike
relation, which for simple fluids reads\cite{KNAW_1914_17_0793}
\begin{equation}
h(r_{12}) = c(r_{12}) +\rho \int c(r_{13})h(r_{32})d{\bf r}_{3}
\label{oz}
\end{equation}
where $h(r)$ is the total correlation function (related to the pair
distribution function, $g$, by $h=g-1$) and $c(r)$ is the direct
correlation function. This equation requires a supplementary
relation, whose general form is\cite{PTP_1960_023_1003}
\begin{equation}
h(r) = \exp\left[-\beta u(r) +h(r)-c(r) + B(r)\right]-1
\label{cr}
\end{equation}
with $\beta=(k_BT)^{-1}$. In Eq.(\ref{cr}) the bridge function, $B(r)$, is a diagrammatic sum of
convolutions of $h(r)$, and must be approximated. The simplest
instance is the hyper-netted chain approximation (HNC), which implies
$B(r)=0$. Interestingly, this approximation predicts the existence of
a LL equilibrium in the square shoulder-square well model studied in
Ref.~\onlinecite{PRE_2002_66_051206} whereas for the A2SRP model only the LV
equilibrium is reproduced. Moreover, in the case of the 2SRP model, it
completely misses the density
anomaly\cite{PRE_2005_72_021501}. Preliminary calculations with a more
elaborate closure such as  the one proposed by Martynov, Sarkisov and
Vompe (MSV)\cite{JCP_1999_110_3961}, show that although it turns out to be
extremely accurate in the determination of the LV equilibria, it fails to
capture the LL transition. However, one observes a second Widom line at
moderate densities, which  seems to indicate an anomaly in
the region where the LL is expected to appear. This  implies
that thermodynamic consistency should play a central role if the LL
transition or the density anomaly is to be found. This is confirmed
by the results  Kumar et al.\cite{PRE_2005_72_021501} for the 2SRP model. 
A fairly successful self consistent approach is the so-called Hybrid
Mean Spherical approximation (HMSA) which smoothly interpolates
between the HNC and the Mean Spherical Approximation (MSA)
closures\cite{PL_1985_108A_277,JCP_1986_84_2336}. The corresponding
closure reads
\begin{eqnarray}
g(r) &=& \exp(-\beta u_r(r))\nonumber\\
& &\times\left(1+\frac{\exp[f(r)(h(r)-c(r)-\beta
  u_a(r))]-1}{f(r)}\right)
\label{hmsa}
\end{eqnarray}
with the interpolating function $f(r) = 1-\exp(-\alpha r)$. The
repulsive component of the interaction given by Eq.(\ref{potr}), and
the attractive component simply given by $u_a(r)=u(r)-u_r(r)$ with
$u(r)$ having been defined in Eq.(\ref{pot}). Note that for the purely repulsive
system, where $u_a=0$, one recovers the Rogers-Young (RY)
closure\cite{PRA_1984_30_999}.  The parameter $\alpha$ is fixed by
requiring consistency between virial and fluctuation theorem
compressibilities. With the virial pressure defined by
\begin{equation}
\beta P^v = \rho -\frac{2\pi}{3}\rho^2\int_\sigma^{d_c} r^3 \frac{\partial
  u_r(r)}{\partial r} g(r) dr +
  \frac{2\pi}{3}\rho^2\sigma^3g(\sigma^+)
\label{Pv}
\end{equation}
and 
\begin{equation}
\beta\kappa_T^{-1}/\rho =\left.\frac{\partial \beta P}{\partial\rho}\right\vert_T = 1 -
4\pi\rho\int_0^\infty r^2 c(r) dr.
\label{chi}
\end{equation}
Consistency could be implemented on a local level by equating
\begin{equation}
 \beta\kappa_T^{-1}/\rho = \left.\frac{\partial \beta P^v}{\partial\rho}\right\vert_T .
\label{LC}
\end{equation}
Alternatively, one might use a global consistency. In this  case, in the absence of
spinodal lines, one can simply enforce the condition
\begin{equation}
\beta P^v(\rho,T) = \beta P^\kappa(\rho,T)
\label{GC}
\end{equation}
with
\begin{equation}
\beta P^\kappa(\rho,T)=\int_0^\rho \left[1 - 4\pi\rho\int_0^\infty r^2
  c(r;\rho',T) dr\right]d\rho'.
\label{GC1}
\end{equation}
In the presence of spinodal lines one might use a mixed thermodynamic
integration path, as suggested by Caccamo et
al.\cite{JCP_2002_117_5072}. In this paper  extensive use is made of the local
consistency (LC) approach. We have also  explored the use of global consistently
(GC) as an alternative to overcome the deficiencies exhibited by the
LC approach. From a practical point of view, in order to calculate the derivative of the
virial pressure in Eq.(\ref{LC}), we shall assume that the parameter $\alpha$ is locally
independent of the density, as is customary. For density independent
potentials this leads to minor discrepancies when comparing the
integrated pressures in Eq. (\ref{GC}), so both LC and GC
approaches yield similar results \cite{JPCM_1993_5_B75}. Here the
situation is somewhat different, 
as will be seen in the proceeding Section. Finally, the right hand side of Eq.(\ref{LC})
is explicitly calculated via
\begin{eqnarray}
 \left.\frac{\partial \beta   P^v}{\partial\rho}\right\vert_T &=& \frac{2\beta
 P^v}{\rho} -1\nonumber\\
 & & -\frac{2\pi}{3}\rho^2\int_\sigma^{d_c} r^3  \left(\frac{\partial
  u_r(r)}{\partial r}\right) \left(\frac{\partial g(r)}{\partial \rho}\right)dr\nonumber\\
 &&+ \frac{2\pi}{3}\rho^2\sigma^3 \left.\frac{\partial g(r)}{\partial \rho}\right\vert_{r=\sigma^+}
\label{dPv}
\end{eqnarray}
where the density derivative of the pair distribution function is
obtained by iterative solution of the integral equation that results
from differentiation of (\ref{oz}) and (\ref{hmsa}) with respect to
density. This procedure, first proposed by
Belloni\cite{JCP_1988_88_5143} is substantially more efficient than
the classical finite differences approach when it comes to evaluating the derivative
of $\beta P^v$. 
In addition to the density derivatives of the pair correlation
function we have also evaluated the corresponding temperature
derivatives in order to calculate the heat
capacity\cite{JCP_2005_122_214504}
\begin{eqnarray}
C_v^* &=& \frac{C_v}{Nk_B}-\frac{3}{2} \nonumber \\
&=&
 2\pi\rho\int\beta u(r)g(r)\left(\beta u(r)-\beta\left.\frac{\partial\omega(r)}{\partial\beta}\right\vert_{\rho}\right)r^2
  dr
\label{cv}
\end{eqnarray}
where $\omega=h-c+B$. Note that the integral equations in terms of the
derivatives of the correlation functions (either with respect to
$\rho$ or $\beta$) are extremely well conditioned and are rapidly
solved, even when starting from ideal gas initial estimates.
\subsection{Simulation algorithms\label{simul}}
Initially the principal aim was to establish whether the A2SRP model 
does indeed exhibit a fluid-fluid equilibrium. Therefore, in order to compute both the liquid-vapour and
the LL equilibria, a procedure \cite{PRE_2005_71_046132} based on the algorithm of Wang and Landau
\cite{PRL_2001_86_2050,PRE_2001_64_056101} was implemented. Subsequently, it was necessary  to compute
the equilibrium of the fluid phases with the low density solid phase.
To that end use was made of thermodynamic integration \cite{frenkel_smit} and the
so-called Gibbs-Duhem integration 
techniques\cite{frenkel_smit,mp_1993_78_1331,jcp_1993_98_4149}, which  took
advantage of the behaviour of the repulsive model in the low temperature limit.
The use of two different methods in independent calculations provides a check on the quality of the results.
\subsubsection{Wang-Landau-type algorithm}
 A more detailed explanation of
the Wang-Landau procedure can be found elsewhere \cite{PRE_2005_71_046132}. Here, we shall
restrict ourselves to outlining its most salient features.
The algorithm fixes 
the   volume and temperature and then samples
over a range of densities ($N$) of the system.
For each distinct density  the Helmholtz energy function $F(N|V,T)$ is calculated.
The procedure resembles, to some extent, well known
grand canonical Monte Carlo methods \cite{frenkel_smit,allen_tildesley},
but using a different (previously estimated) weighting function for the
different values of $N$. 
Once $F(N|V,T)$ is known for different values of $N$,
one is able to  compute the density distribution function for different values of
the chemical potential, $\mu$, and extract the conditions for which the density
fluctuations are maximal. This could indicate the presence of phase
separation and eventually, by
means of finite-size-scaling analysis, one can locate  regions of 
two phase equilibrium and the locus of critical points.\cite{PRE_2005_71_046132}
By following such a procedure one can  easily compute the liquid-vapour (LV) equilibrium
at high temperatures. 
The computation of the liquid-vapour equilibrium was predominantly performed with a simulation
box of volume $V/\sigma^3=1000$ (i.e. box side length of $L^*\equiv L/\sigma=10$). In order to
obtain a precise location of the critical point 
simulation runs were  performed close to the estimated critical point for different
system sizes: $L^*=6,7,8,9,10,11,$ and $12$.
By means of re-weighting techniques and finite-size-scaling analysis, we have
estimated the location of the critical point to be:  $T^*_c=1.487\pm 0.003 $, 
$\rho_c \sigma^3=0.103\pm 0.001$, and $p_c \sigma^3/|W_a| \equiv p_c^* \simeq 0.042$.

At low temperatures the LL equilibrium was found, but
as the temperature was further reduced   we were unable to obtain convergence in
the estimates of the Helmholtz energy function, and the systems eventually formed 
a low density solid  phase. 
Given this situation it was surmised  that a 
triple point, if it exists, between the vapour and the two liquid phases, could not
be stable, and the presence of a stable low density solid  phase is
most likely.
After performing  simulations at several temperatures, with different system
sizes ($0 \le \rho \sigma^3 \le 0.55$, $L^*= 6, 8,  10$), 
it was found that the LL equilibrium  appeared for temperatures below $T^*\simeq 0.38$.
Using the results for $L^*=10$, we can
estimate the critical point for the LL equilibrium at $T_c^* \simeq 0.378 \pm 0.003$, 
$\rho_c \sigma^3 \simeq 0.380 \pm 0.002$ and $(p^*_c/T^*_c) \simeq 0.49 \pm 0.01$.
\subsubsection{Thermodynamic Integration: 2SRP model}
In order to compute the phase equilibria between the low density solid 
and the fluid phases, it is useful to  firstly  consider the repulsive model.
Jagla\cite{JCP_1999_111_8980} computed the stability of different
crystalline structures at $T=0$ for the repulsive model supplemented by a mean
field attraction term, for a ramp model with $d_a/\sigma=1.75$, and found that
the stable crystalline phases at low pressure correspond to the face-centred cubic ({\it fcc}) and 
hexagonal close packed ({\it hcp}) structures. On the other hand, in the limit $T \rightarrow 0$, 
and not too high a pressure
the 2SRP model approaches an effective hard sphere system with diameter $d_a$.
We can, then, expect  the {\it fcc}  structure to be the stable phase in the
range\cite{frenkel_smit}  :
$1.038 (\sigma/d_a)^3 \le \rho \sigma^3 \le \sqrt{2} (\sigma/d_a)^3 $, with the transition
from the low density fluid to the solid at a pressure \cite{frenkel_smit} of 
$\beta p \sigma^3  \simeq 11.6 \: (\sigma/d_a)^3 $.
Assuming that the {\it fcc} is indeed the structure of the low density solid phase, and taking
into account the low temperature limit for the 2SRP model one can use
standard methods to study the equilibrium of this solid with the fluid phase(s).

In order to relate the two ramp models considered in this work, and to
explain the procedures used to compute the phase
diagram of the systems using thermodynamic and Gibbs-Duhem integration, it
is useful to write down a generalised interaction potential, $u_1(r)$ using an additional
`perturbation' parameter, $\lambda$:
\begin{equation}
u_1(r|\lambda) = u_{r}(r) + \lambda   u_a(r).
\label{url}
\end{equation}
%
%
%
When $\lambda=0$, there are two interesting
limits; when $T\rightarrow \infty$  the model  approaches that of a system of hard
spheres of diameter $\sigma$, whereas in the limit $ T \rightarrow 0 $
one once again encounters a hard sphere system, but now having a diameter of $d_a$.
In order to compute the Helmholtz energy for condensed phases of the ramp potential
systems, one can perform thermodynamic integration over the variable $\beta$ and
the parameter $\lambda$. Thus one is able to   compute the difference between the Helmholtz energy 
of a particular state,   and the aforementioned hard sphere systems.

In the canonical ensemble the partition function can be written as:
\begin{equation}
Q_1 = \frac{V^N}{\Lambda^{3N} N!} 
\int d {\bf R}^* 
\exp \left[ - \beta U_1(V,{\bf R}^*,\lambda) \right]
\end{equation}
where $\Lambda$, the  de Broglie thermal wavelength,  depends on the temperature;
 ${\bf R}^*$ represents the reduced coordinates of the $N$ particles:
($\int d{\bf R}^* = 1$), and $U_1$ is the potential energy (given by the
sum of pair interactions $u_1(r|\lambda)$). The Helmholtz 
energy function is given by: $\beta F_1 = - \log Q_1$.
One can then write the Helmholtz energy per particle, $f = F/N$, as:
\begin{equation}
\beta f = \beta f^{id} + \beta f^{ex};
\label{betafidex}
\end{equation}
where the ideal contribution $f^{id}$ can be written as
\begin{equation}
\beta f^{id} =  3 \log \frac{ \Lambda(T) }{\sigma} + \log ( \rho \sigma^3 ) - 1.
\label{betafid}
\end{equation}
and the excess part as
\begin{equation}
\beta f^{ex} = - \frac{1}{N} \log \left\{ 
\int d {\bf R}^* 
\exp \left[ - \beta U_1(V,{\bf R}^*,\lambda) \right]
\right\}.
\end{equation}
In what follows we shall consider the Helmholtz energy as a function of the variables
$N,V,T$, and $\lambda$ (or, equivalently, as $N,\rho,\beta$ and $\lambda$).
We can then compute the derivatives of $\beta f^{ex}  $ with respect to  $\beta$ and 
$\lambda$. I.e.
\begin{equation}
\frac{ \partial (\beta f^{ex}) }{\partial \beta } =  \frac{1}{N}
\langle U \rangle
\end{equation}
and
\begin{equation}
\frac{ \partial (\beta f^{ex}) }{\partial \lambda } =  \frac{1}{N}
\beta  \langle U_a \rangle,
\label{dbfdl}
\end{equation}
where the averaged quantity, $\langle U_a  \rangle$,  is given by:
\begin{equation}
 \langle U_a \rangle =  \left\langle \sum_{i=1}^{N-1}\sum_{j=i+1}^N  u_a(r_{ij}) \right\rangle .
\end{equation}
On the other hand, using the well known thermodynamic relation:
\begin{equation}
\left( \frac{ \partial F }{\partial V} \right)_{NT\lambda} = - p
\end{equation}
one obtains
\begin{equation}
\frac{ \partial (\beta f) }{\partial \rho} = \frac{\beta p}{\rho^2 }.
\end{equation}

For the 2SRP model $(\lambda = 0)$, one can compute the
Helmholtz energy of the low density solid using thermodynamic integration.
As mentioned earlier, in the limit $T\rightarrow 0$ one has an effective
hard sphere system of diameter $d_a$. 
The  phase
diagram of the hard sphere  system \cite{frenkel_smit},
the equation of state for the fluid \cite{jpcm_1997_9_8951,jcp_1969_51_635} 
and the equation of state for the solid \cite{jcp_1998_10_4387} are all well known. 
Thus
\begin{equation}
\beta f^{ex}(\rho,T)
= \beta f_{hs}^{ex}( \rho d_a^3)
- \int_{0}^{T} \frac{U(T_1)}{N k_B T_1^2} \: d T_1.
\label{int_xr}
\end{equation}
where $\beta f_{hs}^{ex}(\rho d^3 )$ is the excess Helmholtz energy per particle in the {\it fcc}-solid
phase and $U/N$ is the excess internal energy per particle.
%
The integrand in (\ref{int_xr}) is  well behaved in the limit $T \rightarrow 0$.
Simulations have been undertaken for $N=500$,  $\rho \sigma^3 = 0.24$
($\rho d_a^3 \simeq 1.2212$),
for various temperatures; $T^*=0.025, ~0.050, ~0.10, ~0.15, ~0.20, \dots, 0.45, ~0.500$.
At $T^*=0.500$ the system melts.
The results for
$U/T^2 $ as a function of $T$ have been fitted using a polynomial. 
Thus we are able to 
obtain a function to compute the Helmholtz energy of the low density solid phase
in the range $0 < T^* < 0.45$ at the {\it reference} reduced density 
of $\rho_{0}^{(s)} \sigma^3 = 0.24$.

In order to acquire data for the calculation of the fluid-low density solid equilibrium
a number of simulations were performed for the low density solid phase, along several isotherms, having densities around 
$\rho_0^{(s)}$.
The low density solid  is stable only within  a small region of the $\rho-T$ phase diagram, as can be seen in Fig. 10.
The pressures in this stability range can be fitted to a polynomial,
\begin{equation}
\beta p_s(T,\rho) = \sum_{i=0} a_{i}^{(s)}(T)  \rho^i.
\end{equation}
We can then compute the Helmholtz energy of the low density solid,
as a function of the density on each isotherm,
using
\begin{equation}
\beta f(\rho_1,T) = \beta f(\rho_0,T) + 
\int_{\rho_0}^{\rho_1}
\frac{\beta p(\rho,T)}{\rho^2} d\rho, 
\label{intfp}
\end{equation}
where $\rho_{0} = \rho_0^{(s)}$, and $\beta p(T,\rho)=\beta p_s(T,\rho)$.

The Helmholtz energy of the low-density fluid
phase is computed using the results of several simulations, having 
$N=500$,  at
several temperatures and densities; typically $\rho \sigma^3 = i \times 0.025$, 
with $i=1,2 \dots, 10$.
In order to guarantee that the samples were indeed within  the fluid phase the simulation
runs were initiated from equilibrated configurations of temperature $T^*  \simeq 2$.
The pressure was fitted to a virial equation of state:
\begin{equation}
\frac{\beta p_g(\rho,T) - \rho}{\rho^2 } = B_2(T) + B_3(T) \rho + B_4(T) \rho^2 + \cdots,
\label{virial}
\end{equation}
which is then used in the calculation of the excess Helmholtz energy of the gas:
\begin{equation}
\beta f^{ex} (\rho, T) =  
\int_{0}^{\rho}  \frac{ \beta p_g(\rho_1,T )-\rho_1 }{\rho_1^2 } \: d \rho_1 
\label{intf1}
\end{equation}

Using (\ref{betafidex},\ref{betafid},\ref{virial}-\ref{intf1}) 
one can compute the Helmholtz energy of the gas phase
as a function of the density.
The Helmholtz energies of the high density fluid can be computed using thermodynamic integration
from the high temperature limit. Performing a number of simulations, with $N=500$,
for different
values of $\beta$ at a fixed reference density, $\rho_{l}$, we have 
\begin{equation}
\beta f^{ex}(\rho_l,\beta) = \beta f_{hs}^{ex}(\rho_l \sigma^3) +  
\int_{0}^{\beta} \frac{U(N,\beta_1,\rho_l) }{N} \: d \beta_1
\label{intbfb0}
\end{equation}
For selected temperatures several simulation runs are performed for
various densities,  and  the results then are used to fit the equation of state  of the liquid branch,
\begin{equation}
\beta p(T,\rho) =  \sum_{i=0} a_i^{(l)}(T) \rho^i.
\label{eosl}
\end{equation}
Similarly, we can compute the Helmholtz energy of the high density fluid at a
given temperature by again performing simulations at several densities, and then fitting
the results of the pressure as a function of the density and using 
(\ref{intfp}) with $\rho_0=\rho_0^{(l)}$ where $\rho_0^{(l)} \sigma^3 = 0.40$.
Once we have the equations of state for the low density fluid, the low density solid, and the high density fluid,
and the corresponding  Helmholtz energies at a certain
reference density for each case, it is straightforward to compute the equilibria
by locating the conditions at which both the chemical potential and pressure
of two phases are equal. 
\subsubsection{Thermodynamic Integration: A2SRP Model}
The Helmholtz energy of the fluid phases can be computed following the same procedures
outlined in the previous subsection. In order to compute the Helmholtz energy of the low density solid phase
one can perform thermodynamic integration starting from the repulsive model and integrating (\ref{dbfdl})
at constant $\beta$, $\rho$, and $N$:
\begin{equation}
\beta f_a(\rho,T)   =
\beta f_{r}(\rho,T) + \frac{\beta}{N} \int_{0}^{1} 
\langle U_a(N,\rho,T,\lambda) \rangle
\: d \lambda,
\label{inttl}
\end{equation}
where the indices $a$ and $r$ indicate  the A2SRP and the 2SRP models respectively.
Such an integration was  carried out using $\lambda= k \times 0.10$ (with $k=0,1,2,\cdots,10$)
with $N=500$ at $\rho \sigma^3 =0.24$ at the temperatures $T^*=0.30$ and $T^*=0.35$.
The integrand $\langle U_a \rangle$ is well behaved in both cases. 
The Helmholtz energy of the A2SRP model low density solid phase at the reference
density $\rho_0^{(s)} \sigma^3 = 0.24$ as a function of $T$ was parametrised via simulation results.
Integration of (\ref{inttl}) at two distinct temperatures
provided a check of the numerical consistency.
As for the 2SRP model various simulations were performed in the region of 
$\rho^{(s)}$ in order to compute the equation of state and the Helmholtz energy
of the low density solid.

Thermodynamic integration was used to study
the equilibrium between
the low density solid and the high density liquid. 
The procedure was analogous  to that used to study  the 
low density solid-high density fluid equilibrium of the 2SRP model . 
To check the results of the LL equilibrium 
thermodynamic integration was performed  at $T^*=0.35$.
In order to do this the Helmholtz energy was calculated at
the `reference' reduced densities of 0.30 and 0.45 using the procedure
derived from (\ref{intbfb0}), and then by performing a simulation
at $T^*=0.35$ at several densities. The results for the equation of state were  fitted 
for both branches, and the conditions
of thermodynamic equilibria were calculated. A good agreement (within the error bars)
was found with the results from the Wang-Landau calculation.
\subsubsection{Gibbs-Duhem Integration: 2SRP Model}
The partition function, $Q(N,p,T,\lambda)$
in the isothermal-isobaric ($NpT$)
ensemble can be written as \cite{frenkel_smit}:
\begin{equation}
Q_{NpT} = \beta p \int d V \: \exp \left[ - \beta p V \right] Q_{NVT}(N,V,T,\lambda).
\end{equation}
The chemical potential, $\mu$
is given by
\begin{equation}
\beta \mu = - \frac{1}{N} \log Q_{NpT}.
\end{equation}
Considering constant $N$, and $\mu = \mu (p,T,\lambda)$ one has:
\begin{equation}
d ( \beta \mu ) = \frac{E}{N} d \beta + \frac{V}{N} d (\beta p) + \frac{\beta U_a}{N} d \lambda.
\end{equation}
where $E$ is the internal energy (including kinetic and potential contributions).
Let us consider two phases, $\alpha$ and $\beta$, 
in conditions of  thermodynamic equilibrium (i.e. equal values
of $T$, $p$ and $\mu$); if one makes a differential 
change in some of the variables $\beta$, $(\beta p)$ and
$\lambda$, thermodynamic equilibrium will be preserved if:
\begin{equation}
d [ \Delta (\beta\mu)] = 0 =  \Delta \bar{U} d \beta  +
\Delta \bar{V} d (\beta p)  +
\Delta  \bar{U}_a  d  \lambda.
\label{ddbmu}
\end{equation}
where $\Delta X$ expresses the difference between the values of the property $X$ in
both phases, and ${\bar X} \equiv X/N$.
In (\ref{ddbmu}) we have considered equal values of the kinetic energy per particle in both
phases (classical statistics).

In order to compute the fluid-low density solid equilibrium of the 2SRP model we  
make use of the Gibbs-Duhem integration  scheme.
As the starting point of the integration we consider the fluid-solid
equilibrium of the effective hard sphere system, with diameter $d_a$ at $T=0$.
We have integrated (\ref{ddbmu}) for $\lambda=0$. 
After a number of short exploratory simulation runs, the Gibbs-Duhem integration
 computation for the 2SRP model
was performed in three subsequent stages:
i) low density fluid-low density solid equilibrium at low temperature,
ii) fluid-low density solid at `higher' temperatures, and iii) low density solid-high 
density fluid at low temperatures.
For the first stage equation  (\ref{ddbmu}) was modified to obtain (in reduced units)
the finite interval numerical approach:
\begin{equation}
\delta   \left(\frac{p}{T}\right) \simeq \frac{ \Delta \bar{U}}{T^2 \Delta \bar{V}} \delta T.
\label{deltabp}
\end{equation}
The calculation of
$\bar{U}$ and ${\bar V}$ for both phases at the estimated coexistence conditions
was performed by Monte Carlo simulations with $N=500$, in the $NpT$ ensemble. 
At $T^*=0$ the coexistence of the fluid
and the solid phases occurs \cite{frenkel_smit} at 
 $(p^*/T^*) \simeq 2.286$ (i.e. $\simeq 11.6 \times (\sigma/d_a)^3$).
A number of  simulations ($T^*=0.01, 0.02$ with $(p^*/T^*) \simeq 2.29$)
were used to estimate the initial values
of the slope $d(p^*/T^*)/dT^ *$. 
The integration was then carried  out from $T^*=0.01$ to $T^*=0.40$ with a 
step of $\delta T^*=0.01$,

In the second stage of the integration the coexistence temperature reaches
a maximum, thus the independent variable in the integration was substituted for
\begin{equation}
\delta T \simeq \frac{T^2 \Delta {\bar V} }{\Delta {\bar U} }  \delta (p/T).
\label{deltat}
\end{equation}

This integration was started from the pressure at which the temperature
reaches $T^*=0.40$ (i.e. $(p^*/T^*) \simeq 3.65$)
using an integration step of $\delta (p^*/T^*) = 0.05 $. Initially, the
temperature increases with $p$ until the coexistence line reaches
a maximum temperature;  at this point the density of both phases
is equal: ($T=0.422\pm0.002$, $p^*/T^*=4.65\pm 0.05$, $\rho \sigma^3 = 0.260\pm 0.002$).
For higher pressures the fluid density becomes higher than
the low density solid density, and  the temperature decreases with $p/T$. 
The third stage is initiated upon reaching 
$T^*=0.40$; which happens at $(p^*/T^*) \simeq 6.33$. This involves reverting 
to
the integration scheme given by 
(\ref{deltabp}), this time using $\delta T^* = -0.01$.
\subsubsection{Gibbs-Duhem Integration: A2SRP Model}
In order to compute the fluid-low density solid equilibria of the A2SRP 
system, this system is connected to the fluid-low density solid equilibria of the 2SRP model,
by tuning the
perturbation parameter $\lambda$ from $\lambda=0$ to $\lambda=1$, while maintaining
the conditions of thermodynamic equilibrium, i.e. those of Eq. (\ref{ddbmu}).
For this system this connection stems from the initial state:
$\lambda=0,T^*=0.34, (p^*/T^*) = 3.05$, which corresponds to the
equilibrium between the low density solid  with a lower density fluid.
This was done by keeping $T$ constant and integrating numerically the equilibrium
pressure as a function of $\lambda$, i.e.:
\begin{equation}
\delta (p^*/T^*) \simeq - \frac{ \Delta {\bar U}_a }{\Delta {\bar V} } \delta \lambda.
\end{equation}
using $\delta \lambda = 0.025$. The final
point was found to be $(\lambda=1,p^*/T^* = -0.324)$, corresponding to
a  equilibrium between a low density solid with density $\rho^*=0.248$ and a low
density liquid with $\rho^*=0.239$. Given the negative
sign of the pressure the equilibrium found is metastable.
From this point we can retake
the Gibbs-Duhem integration (now for the attractive model, i.e. setting $\lambda=1$); 
using (\ref{deltat}), with an integration step of
$\delta (p^*/T^*) = 0.01$. As with the repulsive model  a temperature maximum
was found at $T_{M}^* \simeq 0.345$, $\rho \sigma^3 \simeq 0.255$.
However,
in this case such a maximum seems to be metastable:
$(p^*/T^*) \simeq -0.01$. The solid-fluid equilibrium 
becomes thermodynamically stable at $p^*/T^* \approx 10^{-4}$
(where this coexistence line crosses the liquid-vapour equilibrium line).
At higher pressures and lower temperatures this low density liquid-low density solid coexistence line will 
eventually coincide with the LL
equilibrium line at a triple point. 
Beyond this a stable low density solid-high density liquid equilibrium will appear.


In order to obtain a precise estimate of the conditions for which the two
liquid phases and the low density solid are in equilibria (i.e. the location of the triple point),
Gibbs-Duhem integration was performed for the LL  equilibrium. 
Both thermodynamic integration and the Wang-Landau procedure were
used to calculate the LL  equilibrium, at  $T^*=0.35$.
Both of the results agreed to within the error bars: ($\beta p \sigma^3 \simeq 0.465$,
with reduced densities of the liquid phase: $\rho_1\sigma^3 \simeq 0.30$, and
$\rho_2\sigma^3 \simeq 0.46$). 
From this initial point the Gibbs-Duhem integrations of the
phase equilibrium were carried out using (\ref{deltabp}) with $\delta T^*=0.0025$.
The LL equilibrium line met the low density liquid-low density solid equilibrium at $T^* \simeq 0.331$, $(p^*/T^*) \simeq 0.437$,
with the three phases having reduced densities: $\rho_{\rm (low~density~solid)}\sigma^3 \simeq 0.263$,
$\rho_{\rm (low~density~liq.)}\sigma^3 \simeq 0.293$ and $\rho_{\rm (high~density~liq.)}\sigma^3 \simeq 0.48$ .
Gibbs-Duhem integration was then used to calculate the 
low density solid-high density liquid equilibrium, starting from this triple point,
using (\ref{deltabp}) with $\delta T^* =0.01$
\subsubsection{Vapour-solid equilibrium}
The computation of the density of the low density solid at equilibrium with the vapour phase
for the A2SRP model
is straightforward. In practice, due to
the low pressure at which this equilibrium occurs, the $NpT$ simulations of the low density solid 
are undertaken at 
$\beta p \simeq 0$, again with $N=500$. This immediately yields an
accurate estimate of the density of the solid in equilibrium with the
vapour phase.
\subsubsection{Details of the Gibbs-Duhem integration scheme}
A simple predictor-corrector scheme was used to build up the coexistence lines.
Generically one seeks the solution of:
\begin{equation}
\frac{dy}{dx} = f(x,y).
\end{equation}
from some initial condition $x_0,y_0$. In the present case the function $f(x,y)$
has to be computed via a pair of computer simulations. This result will
be affected by statistical error.
Using a set of discrete values $x_i = x_0 + i \cdot h$, the following scheme was used:
\begin{equation}
y^{(p)}_{i+1} = y_i + \frac{1}{2} \left( 3 f_i - f_{i-1} \right) h
\label{gdieq1}
\end{equation}
\begin{equation}
y_{i+1} = y_i + \frac{1}{2} \left( f_i + f_{i+1} \right).
\label{gdieq2}
\end{equation}
with:
\begin{equation}
f_{i} = f(x_{i},y^{(p)}_{i} )
\label{gdieq3}
\end{equation}
where (\ref{gdieq1}) and (\ref{gdieq2}) are respectively the predictor
and corrector steps. 
This simple algorithm is both accurate and robust.
\section{Results}
\begin{figure}
\includegraphics[width=8cm,clip]{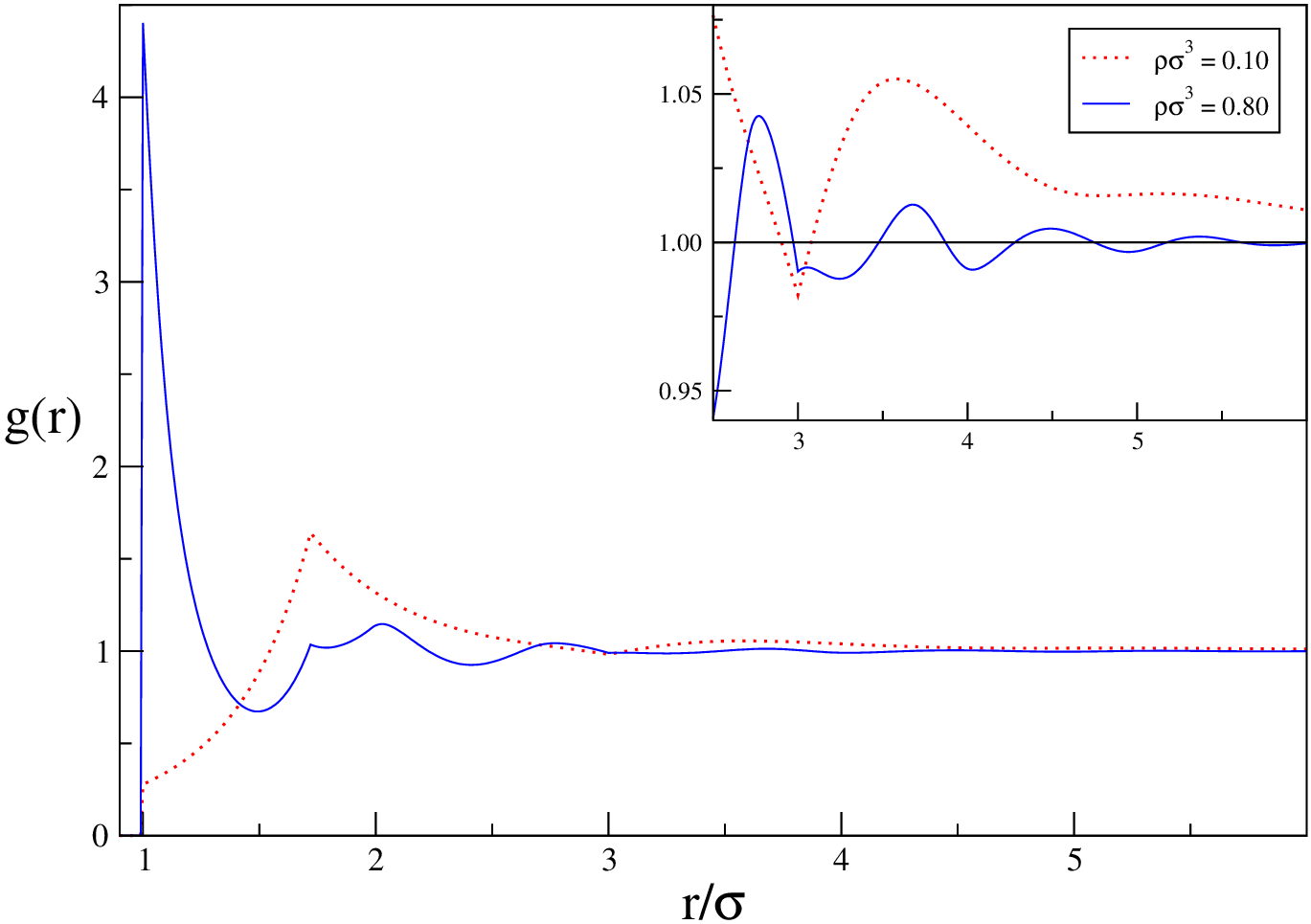}
\caption{(Colour online) Pair distribution function for the A2SRP fluid
  at low ($\rho \sigma^3 = 0.10$) and high ($\rho \sigma^3 = 0.80$) densities at $T^*=1.8$. Inset: magnified view of the region $r/\sigma$ from 2.5 to 6. }
\label{fig2}
\end{figure}
In order to illustrate the two competing LPS that the A2SRP model system
exhibits at low and high densities (or low and high pressures
respectively) in Figure \ref{fig2} we have presented the pair
distribution functions for two representative states of the A2SRP
model, obtained from the
HMSA integral equation. The high density
state is close to a hard sphere fluid of diameter $\sigma$, whereas
the low density $g(r)$ corresponds to a fluid of soft particles of
diameter $d_a = 1.72\sigma$. The purely repulsive system leads to similar
results, with only minor differences, due to the lack of dispersive forces (the
sharp minima at $3\sigma$ are absent). It is clear that the LL
transition will result from the `chemical equilibrium' between two
essentially different `fluids' that stem from the same interaction. 
\begin{figure}
\includegraphics[width=8cm,clip]{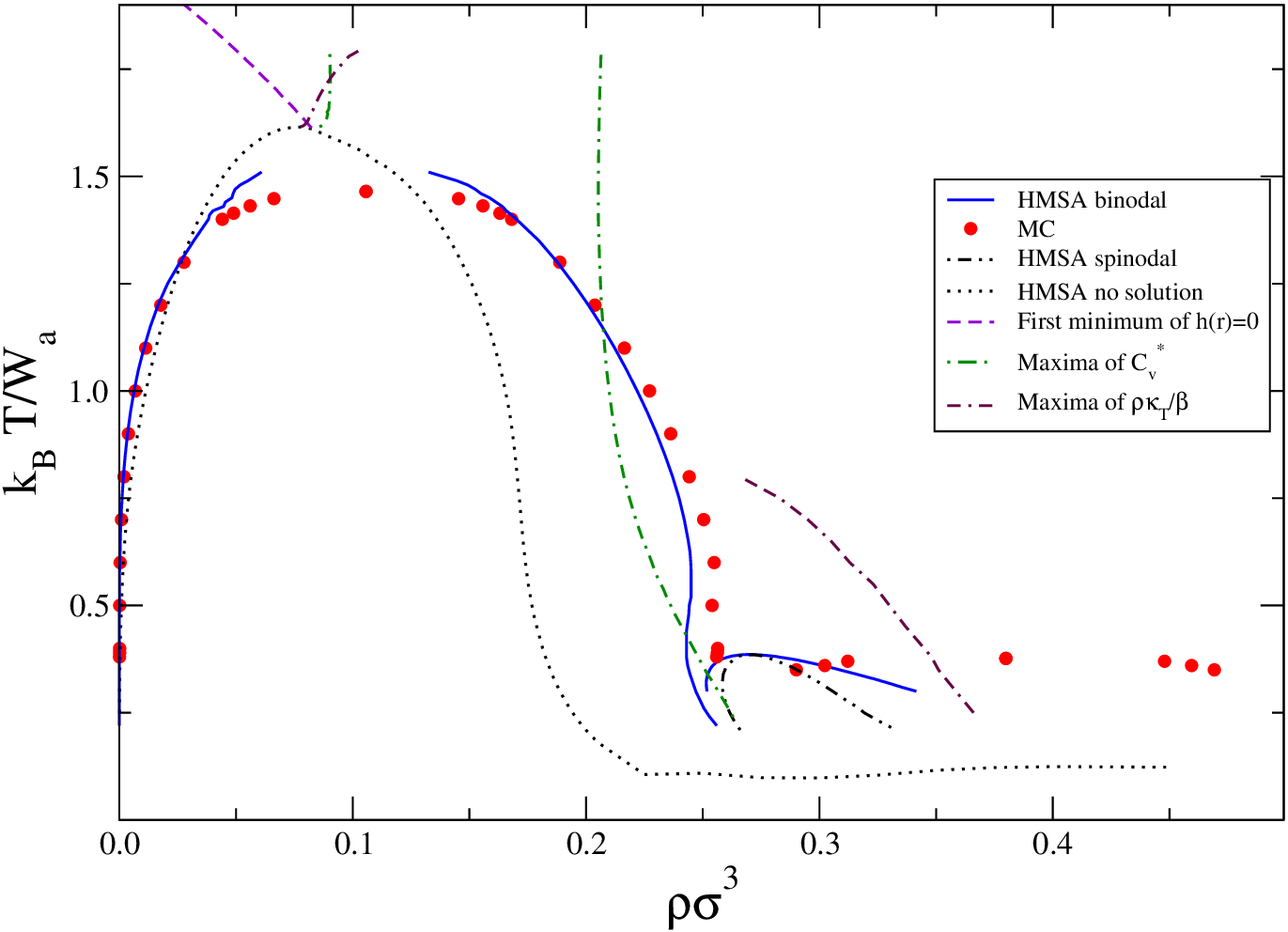}
\caption{(Colour online) Phase diagram of the A2SRP model as obtained
  from the HMSA and Wang-Landau Monte Carlo simulations. HMSA results for the loci of maxima of
  $C_v^*$ (the Widom line), isothermal compressibility
  ($\rho\kappa_T/\beta$) and the set of thermodynamic states for
  which the first non trivial minimum of $h(r)$
  vanishes. Additionally,  the boundaries of the
  non-solution region of the integral equation and the thermodynamic
  spinodal of the LL transition are also depicted.}
\label{fig3}
\end{figure}
\subsection{Liquid-vapour and liquid-liquid equilibria}
\begin{figure}
\includegraphics[width=8cm,clip]{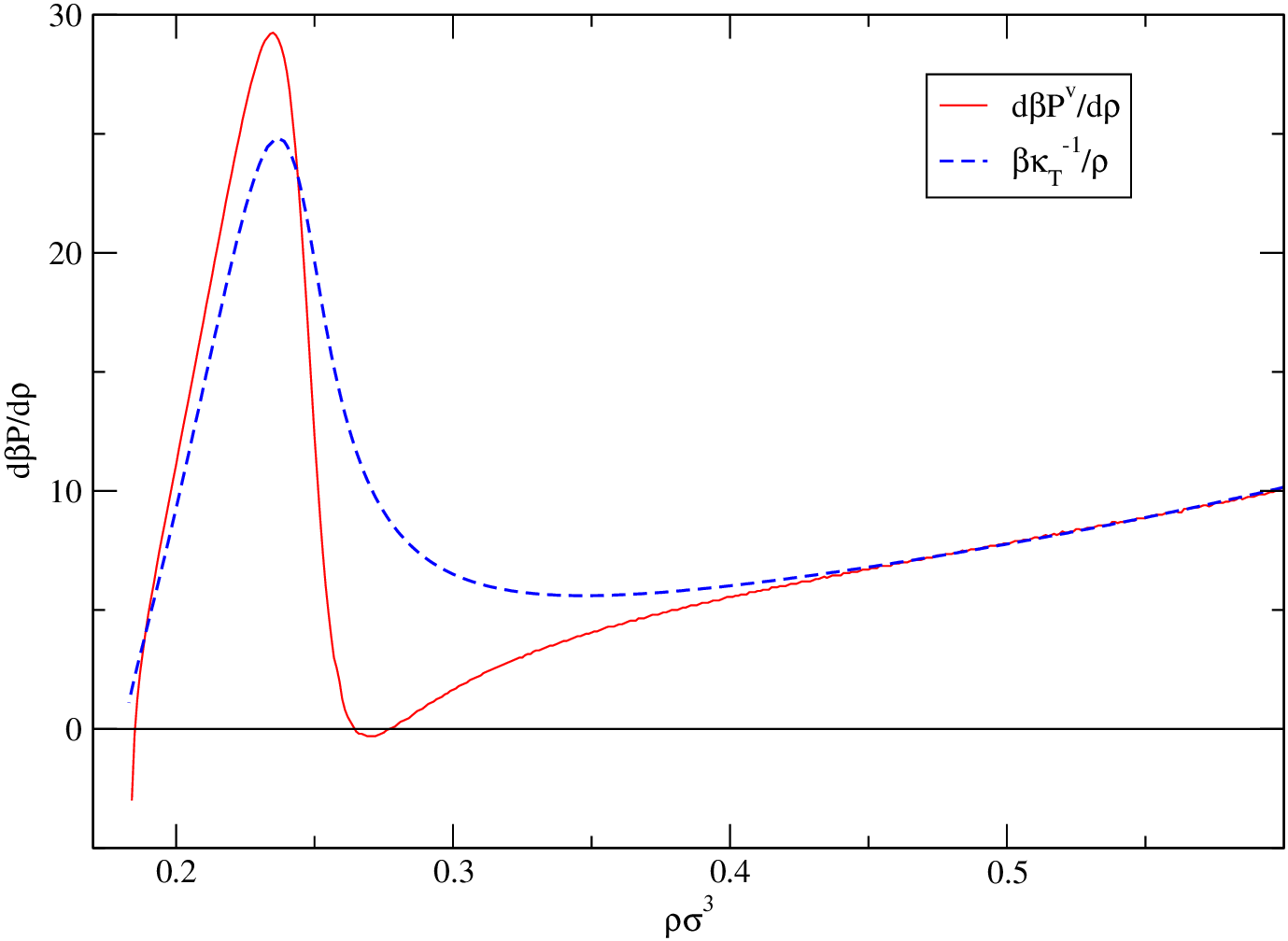}
\caption{(Colour online)  Inverse virial and fluctuation theorem
  compressibilities (solid and dashed curves respectively) as
  calculated using the HMSA integral equation at $T^*=0.38$ for the attractive model. Note how the consistency
  degrades when approaching the LL transition region (see Fig.\protect\ref{fig3}) to the point
  that the virial compressibility crosses a thermodynamic LL spinodal
  before hitting the LV spinodal, whereas the fluctuation theorem
  compressibility only detects the presence of the LV spinodal.}
\label{fig4}
\end{figure}

In order to calculate the phase diagram using the HMSA integral
equation, one has to resort to thermodynamic integration. Although direct
closed formulae for the chemical potential are available in the work
of Belloni\cite{JCP_1988_88_5143}, it has been the authors  experience that better
results are obtained when using thermodynamic integration. For a
given state point, at density $\rho$ (on the right hand side of the phase
diagram) and a subcritical inverse
temperature $\beta$, one can obtain the free energy per particle along a mixed path
of the form
\begin{eqnarray}
f(\rho,\beta) &= \frac{1}{\beta_0}\int_0^\rho
(\beta p(\rho',\beta_0)-\rho)/\rho'^2 d\rho'\nonumber\\
&  + \frac{1}{\beta}\int_{\beta_0}^\beta
U(\rho,\beta')/Nd\beta'+\log\rho\Lambda^3 -1
\label{frho}
\end{eqnarray}
where $\beta_0$ is a supercritical inverse temperatureWith this
expression one is able to  evaluate the chemical potential by using the relation 
\[ \beta\mu = \beta f + \frac{\beta p}{\rho}. \]
The phase equilibria can be determined by equating pressures and
chemical potentials for both the gas and liquid phases, and for both the low density
liquid and high density liquid phases. 

The phase diagram thus obtained is plotted in Figure \ref{fig3}
together with the corresponding Monte Carlo
estimates. Additionally, the curves corresponding to the loci of maxima of
  $C_v^*$ (the Widom line), maxima of the isothermal compressibility
  ($\rho\kappa_T/\beta$) and the set of thermodynamic states for
  which the first non trivial minimum of $h(r)$
  vanishes are presented. This latter quantity separates supercritical
  states with gas-like local order from those with liquid-like
  order\cite{JCP_2003_119_373}. Interestingly, as found in the
  Lennard-Jones system with a completely different
  closure\cite{JCP_2003_119_373}, these three
  singular lines are seen to converge towards the LV critical point. Although this
  should be expected from the locus of heat capacity and
  compressibility maxima, there is no apparent fundamental reason why
  this should also happen for the line of states for which the first
  minimum of $h(r)=0$. Moreover, we have
  found that similar results (with different locations for the
  critical point) are obtained using different closures, such as the
  HNC, Reference HNC or the MSV. Finally, one sees that the upper part
  of the binodal line is broken, since convergent solutions cease to
  exist before the critical point is reached. In fact, a small portion
  of the high temperature LV binodal has been obtained by
  extrapolation of 
  the chemical potentials and pressures at lower vapour densities
  (for this reason the no-solution  and  the binodal
  curves cross). This is a well known limitation of a number of integral
  equation approaches, encountered even when treating  the simple Lennard-Jones
  fluid\cite{MP_1989_68_87}. 
\begin{figure}
\includegraphics[width=8cm,clip]{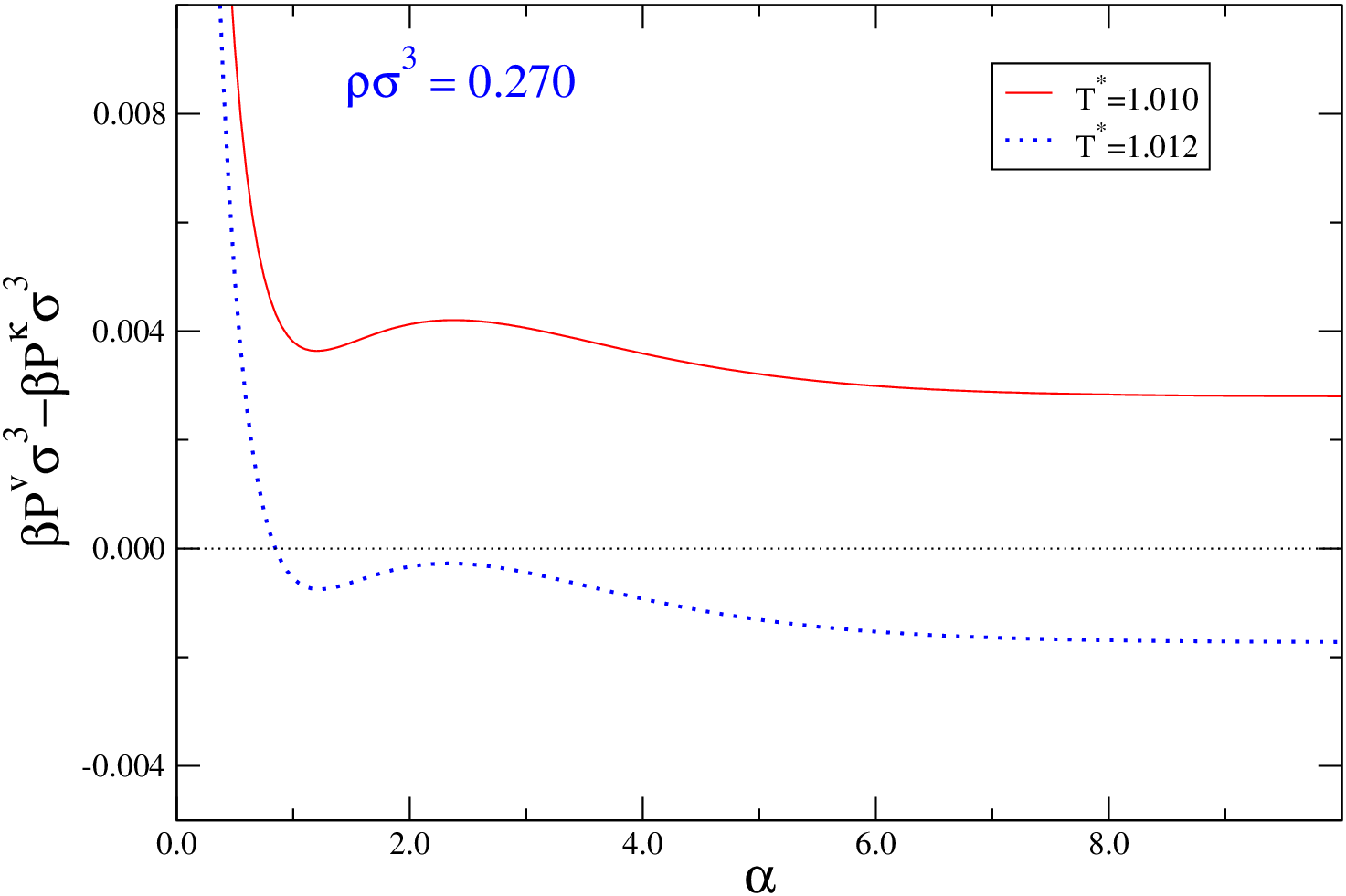}
\caption{(Colour online)  Residual inconsistency between virial and
  fluctuation theorem pressure as a function of the mixing parameter
  $\alpha$ in the HMSA closure (Eq. \protect\ref{hmsa}) for the 2SRP model.}
\label{fig5}
\end{figure}

As for the LL transition, in Figure \ref{fig3} the HMSA
  binodal results obtained from the thermodynamic integration, its
  corresponding thermodynamic spinodal curve and the  Wang-Landau Monte Carlo binodal
  estimates are presented. A second
  Widom line that crosses the LL binodal in the neighbourhood of the HMSA
  critical point is also found. As well as this, a second line of isothermal compressibility maxima
  seems to indicate the presence of a LL transition, approaching the
  simulation LL critical point. Note, however, that in this case the no-solution line
  does not provide any clue as to the presence of a phase
  separation. In contrast to the HMSA behaviour in the vicinity of the LV
  critical point, now the curves of $C_v^*$ and $\rho\kappa_T/\beta$ maxima do
  not converge to a common estimate of the critical point. This reveals a substantial
  inconsistency between the virial and the fluctuation theorem
  thermodynamics. This is best illustrated in Figure \ref{fig4} where
  the virial and fluctuation theorem
  compressibilities are plotted. Here it is observed that the local consistency
  criterion of Eq.(\ref{LC}) leads to a good agreement for densities
  $\rho\sigma^3 > 0.4$, but a considerable inconsistency shows up at
  lower densities, corresponding  to the LL transition (see Fig.\ref{fig3}). The assumption of a density independent $\alpha$
  parameter in the HMSA closure (Eq. \ref{hmsa}) lies at the root of this
  failure. Thus, whereas the virial pressure predicts a LL transition,
  no diverging correlations appear near the thermodynamic
  spinodal, and the equation has solutions throughout the two
  phase region. This is in marked contrast to the results for the LV
  transition. On the other hand, in Fig.\ref{fig4} the large values of
  $(\partial\beta P/\partial\rho)_T$ in the region
  $0.2<\rho\sigma^3<0.3$ show that the fluid is almost incompressible,
  which is somewhat surprising, especially  at these relatively low densities. This
  indicates that a solid phase (or a glassy state) may well be about to appear.
We have attempted to explore alternatives  to this breakdown in the LC
  approach. After an unsuccessful attempt to incorporate a linear
  density dependence on the $\alpha$ parameter in the HMSA
  interpolating function, in conjunction with a
  two parameter optimisation, we focused on the GC alternative
  expressed in Eq.(\ref{GC}). This can be easily implemented
  using the strategy proposed by Caccamo et
  al.\cite{JCP_2002_117_5072}. To make matters simpler, 
  the GC HMSA calculations were restricted to the purely repulsive 2SRP model. At relatively high
  temperatures ($T^* \approx 2$) the procedure worked well and lead to
  results slightly better that those of the LC HMSA. However, as the
  temperature is lowered, one finds that the optimisation loop
  diverges even for relatively low densities. The reason behind this
  divergence is illustrated in Fig.\ref{fig5}. It was  observed that for two
  neighbouring states, a slight decrease in temperature upwardly displaces the
  residual inconsistency curve. As a consequence of this, when looking
  for consistency, the minima are now found to be greater than zero, thus
  the numerical iterations will lead to either a divergence or
  result in an oscillating behaviour. This also explains why the alternative
  procedure based on the implementation of a density dependent
  $\alpha$ failed as well. A completely parallel situation is  to
  be found for the attractive potential model. It is now clear that a
  more accurate treatment of this type of soft core models requires
  not only the implementation of GC conditions on the pressure, but
  also the use of more flexible closure relations.

\begin{figure}
\includegraphics[width=8cm,clip]{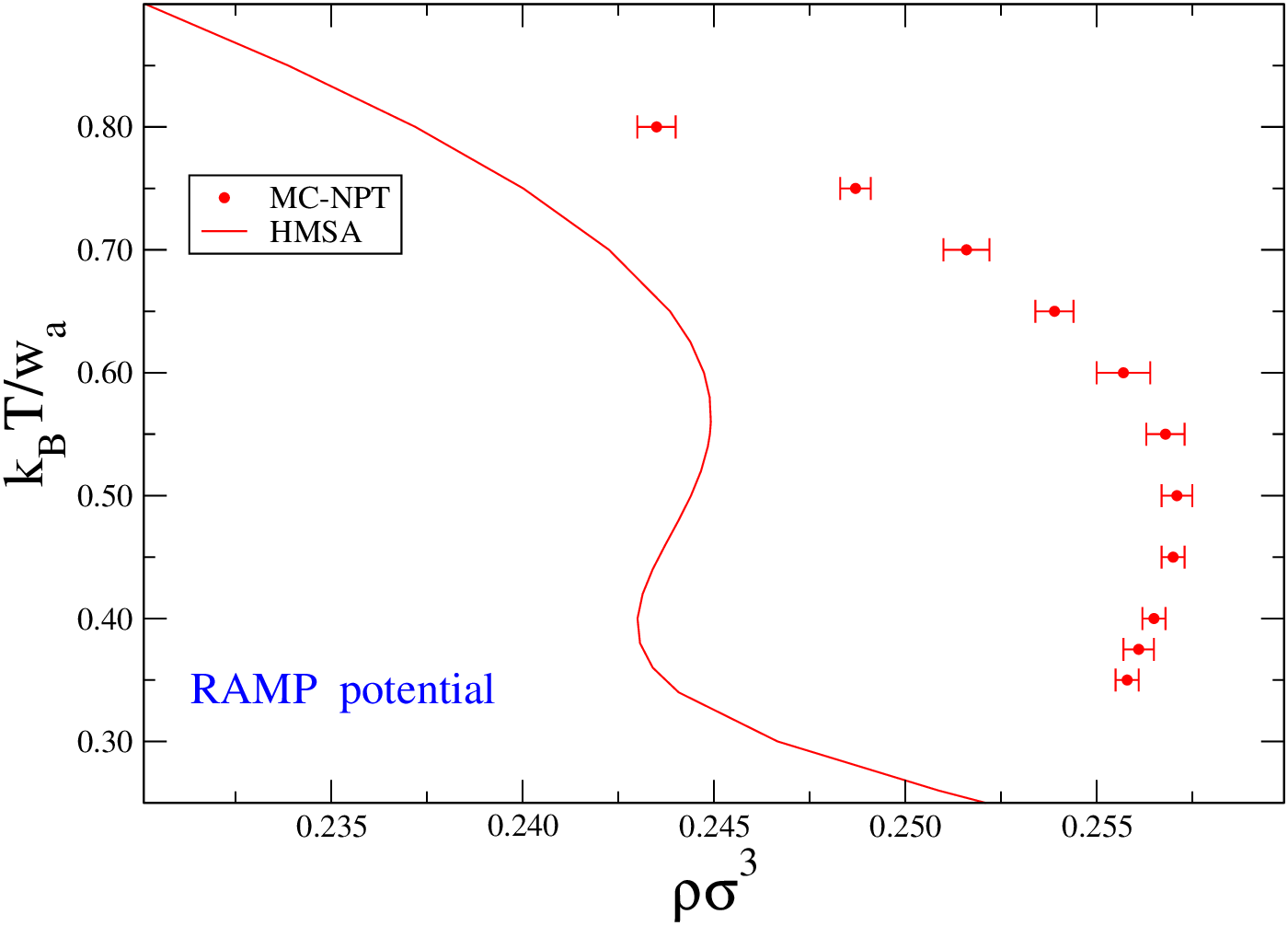}
\caption{(Colour online)  Zero pressure isobar calculated using the HMSA
  scheme and also by means of NPT MC simulation, for the A2SRP model.}
\label{fig6}
\end{figure}

Finally, a few words regarding the comparison of the integral equation
estimates with the simulation results. As far as the LV transition is
concerned, the HMSA predictions are satisfactory. The re-entrant
behaviour that is clearly seen on the high density side of HMSA LV
curve is also found in the MC estimates, and can be more clearly
appreciated in the zero pressure isobar depicted in
Fig.~\ref{fig6}. Here we see that the errors in the density do not exceed
5 per cent. This partly reentrant behaviour is a characteristic indication of
the proximity of a triple point, and has also been found in various water
models \cite{PRL_2004_92_255701,JCP_2005_123_044515}. On the other hand, whereas the LL critical temperature is
reasonably well captured by the theory, the critical density is substantially
underestimated, and the global shape of the LL equilibrium curve is
not well reproduced. This can certainly be ascribed to the poor
consistency of this HMSA approach in this region. Note, however, that
this is the only integral equation theory, of those that we have checked, that
captures the presence of the LL transition.

\begin{figure}
\includegraphics[width=8cm,clip]{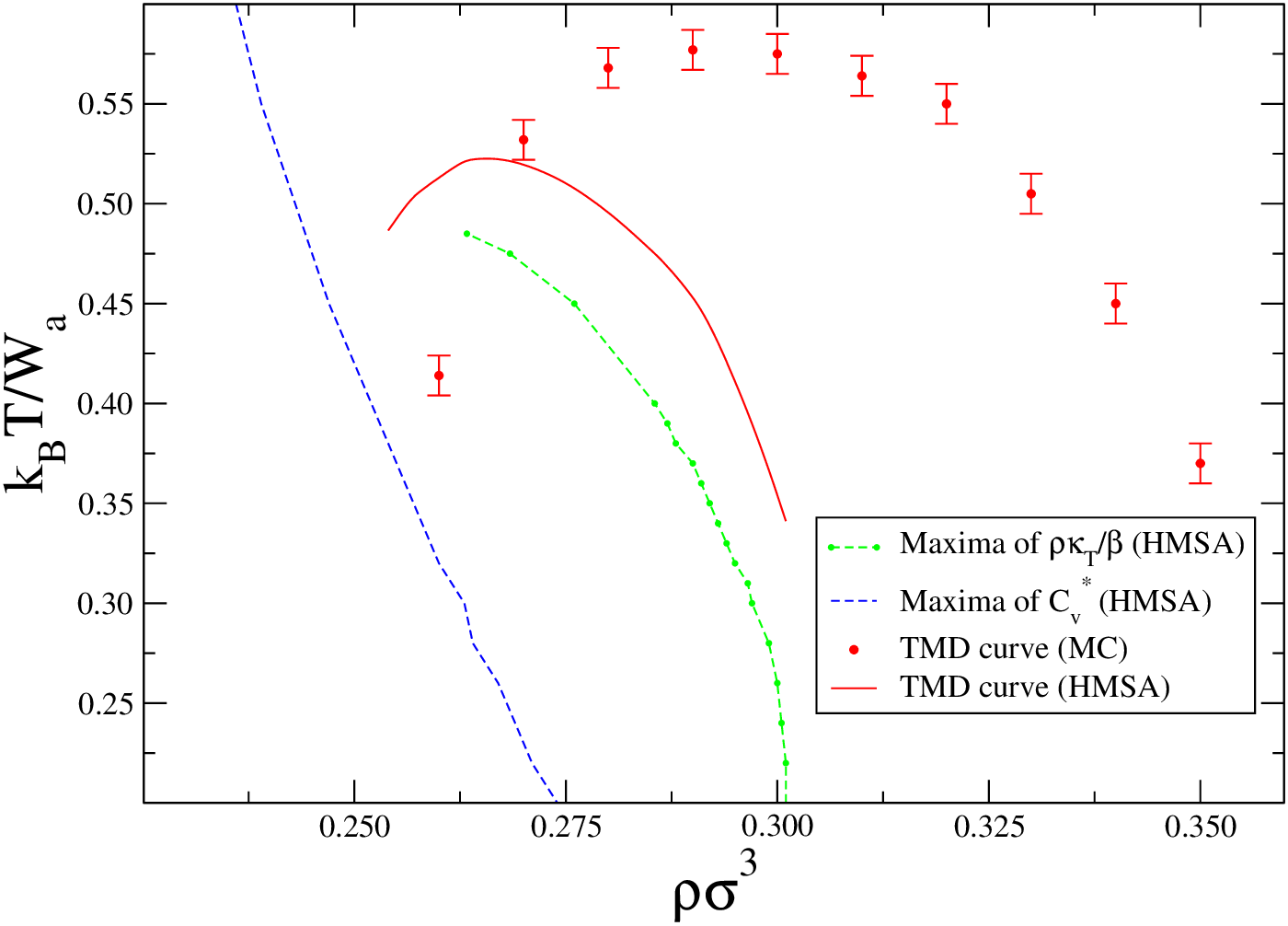}
\caption{(Colour online)  TMD curves as
  determined by HMSA calculations and NPT MC simulation for the purely repulsive model. Curves of maxima in
  the isothermal compressibility and heat capacity (Widom's line) calculated  using the HMSA closure.}
\label{fig7}
\end{figure}

When the attractive interactions are switched off, the LV transition
disappears. The HMSA calculations yield lines of compressibility and heat
capacity maxima that could be interpreted as indications of a possible LL
transition (see Figure \ref{fig7}). These lines appear in approximately  the
same locations as in the attractive model (see Fig.\ref{fig3}), which indicates
that this feature is entirely due to the presence of the soft repulsive ramp
in the interaction. The possibility of a LL transition in this model has  
already been speculated upon by Kumar et al.\cite{PRE_2005_72_021501} on the basis of
the low temperature asymptotic behaviour of a series of isochores. The same
conclusion might well be drawn from this work; however, once again calculations at lower
temperatures are hindered by lack of convergence. It shall be seen in the next
section that the fluid-solid equilibrium preempts the LL phase separation.

As in
Ref.~\onlinecite{PRE_2005_72_021501}, we are able to calculate the TMD curves (lines
for which $(\partial T/\partial\rho)_P = 0)$ from the minima of the pressure
along isochoric curves (i.e. $(\partial P/\partial T)_\rho = 0$). These results
are plotted in Fig.\ref{fig7} and are compared with canonical MC estimates. The
HMSA is qualitatively correct, with errors not exceeding 15 per
cent. Nonetheless, once more the inconsistency of this LC HMSA is
encountered in the results. From a simple thermodynamic analysis it is
known that the line of maxima of the isothermal
compressibility must cross the TMD curve at its maximum temperature \cite{PRE_1996_53_6144}. In
Fig.\ref{fig7} one can observe that the curve of compressibility maxima disappears
before reaching the TMD. Moreover, an extrapolation would locate the crossing
at approximately  $\rho\sigma^3=0.255$ and $T^*=0.496$, meanwhile the maximum of the TMD
appears at $\rho\sigma^3=0.265$ and $T^*=0.523$. Whereas this can be accepted
as being qualitatively correct, we find once more that global consistency must be
enforced if quantitative predictions are to be obtained.

\subsection{Solid-fluid equilibria}
\begin{figure}
\includegraphics[width=8cm,clip]{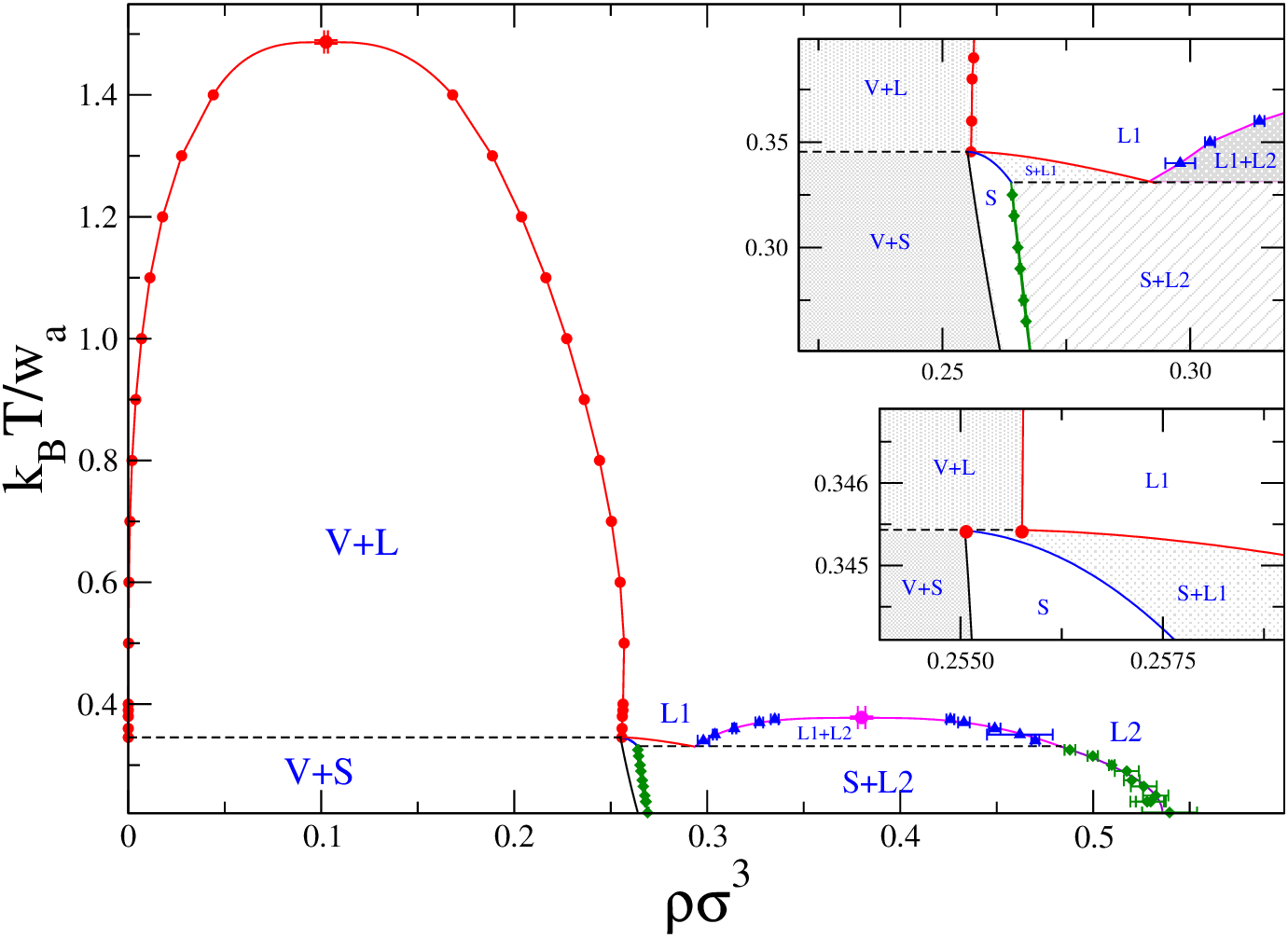}
\caption{(Colour online)  T-$\rho$ phase diagram of the attractive A2SRP model as
  obtained from computer simulation (circles, LV Wang-Landau MC estimates),
  (triangles, LL Wang-Landau MC estimates), (diamonds, MC NVT thermodynamic
  integration), (curves, Gibbs-Duhem integration) (VS equilibrium line,
  zero pressure NPT MC).
  The line connecting the liquid-vapour equilibrium points is drawn as a guide for the eye.}
\label{fig8}
\end{figure}

\begin{figure}[b]
\includegraphics[width=8cm,clip]{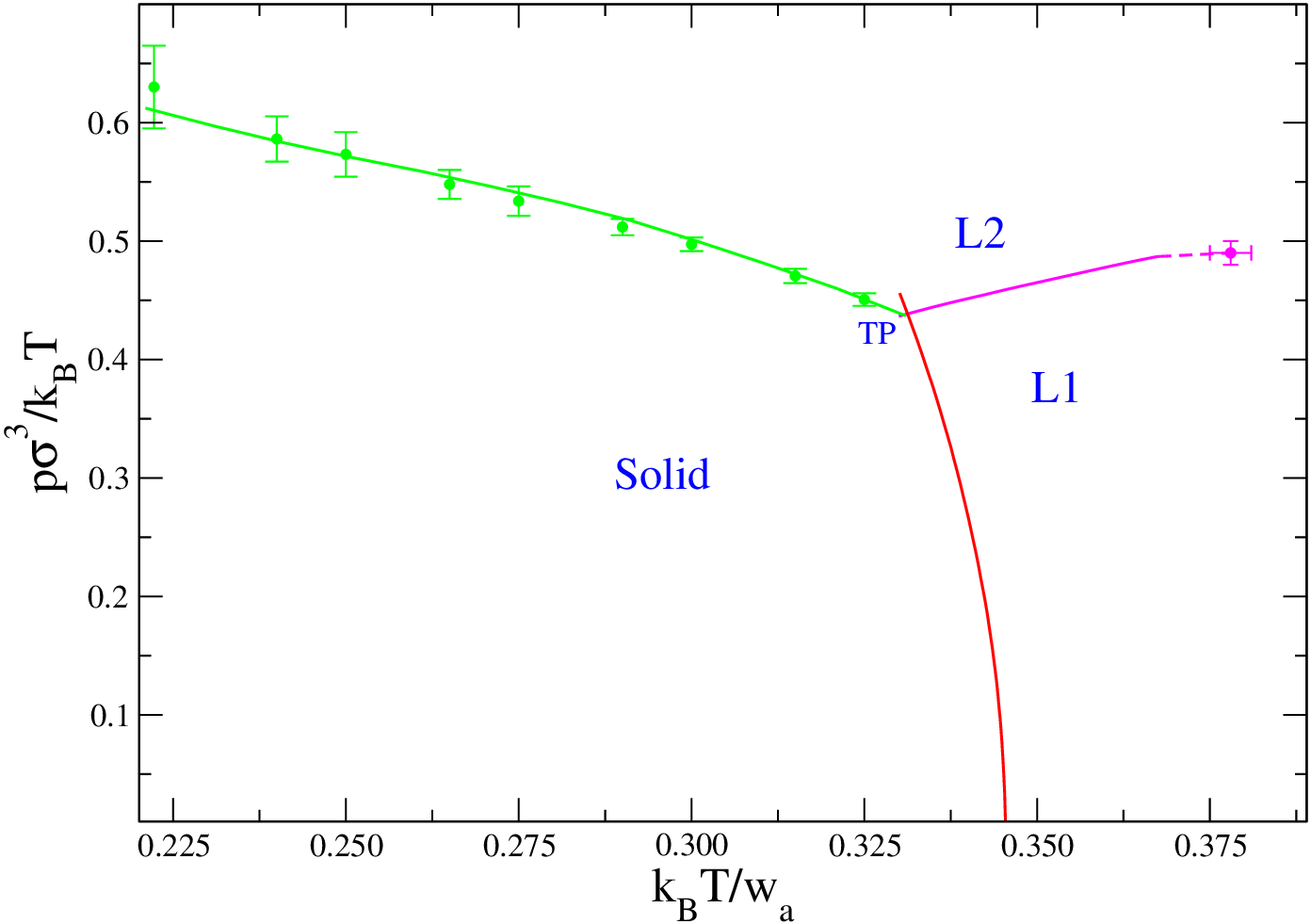}
\caption{(Colour online)  P-T phase diagram of the attractive A2SRP model as
  obtained from computer simulation. Points: thermodynamic integration, 
lines: Gibbs-Duhem integration. Legend as in Fig.\protect\ref{fig8}.}
\label{fig9}
\end{figure}

As explained in Subsection \ref{simul}, the complete phase diagram, including
the {\it fcc} solid phase which appears at moderate density\cite{JCP_1999_111_8980},
is computed by means of a combination of Wang-Landau Monte Carlo simulation
(for the determination of LV and LL equilibrium curves), NPT MC zero pressure
calculations to determine the VS equilibrium curves, standard thermodynamic
integration and Gibbs-Duhem integration to evaluate the various fluid-solid
equilibria. A detailed phase diagram is presented in Figs. \ref{fig8} and
\ref{fig9}. Note that for densities above $\rho\sigma^3=0.5$ other solid
phases are possible. From Jagla's work\cite{JCP_1999_111_8980} (in which
the relative stability of the zero temperature solid phases  is explored) one
can infer that at somewhat higher densities rhombohedric, cubic and hexagonal ({\it hcp})
phases could be expected. At still higher densities, for which the hard cores
play the leading role, again the {\it fcc} and {\it hcp} phases will be the most stable
structures. In any case, from Figure \ref{fig8} one can already see a fairly
complex phase diagram resulting from the coexistence of a vapour phase, two
liquid phases and the {\it fcc} solid. Inset within Figure \ref{fig8} we have enlarged
the area of multiple coexistence, where one observes the presence of three
triple points, namely two neighbouring triple points (lower inset) in which one
liquid phase (denoted by L1), the vapour and the solid coexist, and a third triple point
at lower temperature in which the two liquid phases (L1 and L2) coexist with
the {\it fcc} solid. This latter triple point is also depicted in Fig.\ref{fig9} in a detail
of the $P-T$ phase diagram.

\begin{figure}
\includegraphics[width=8cm,clip]{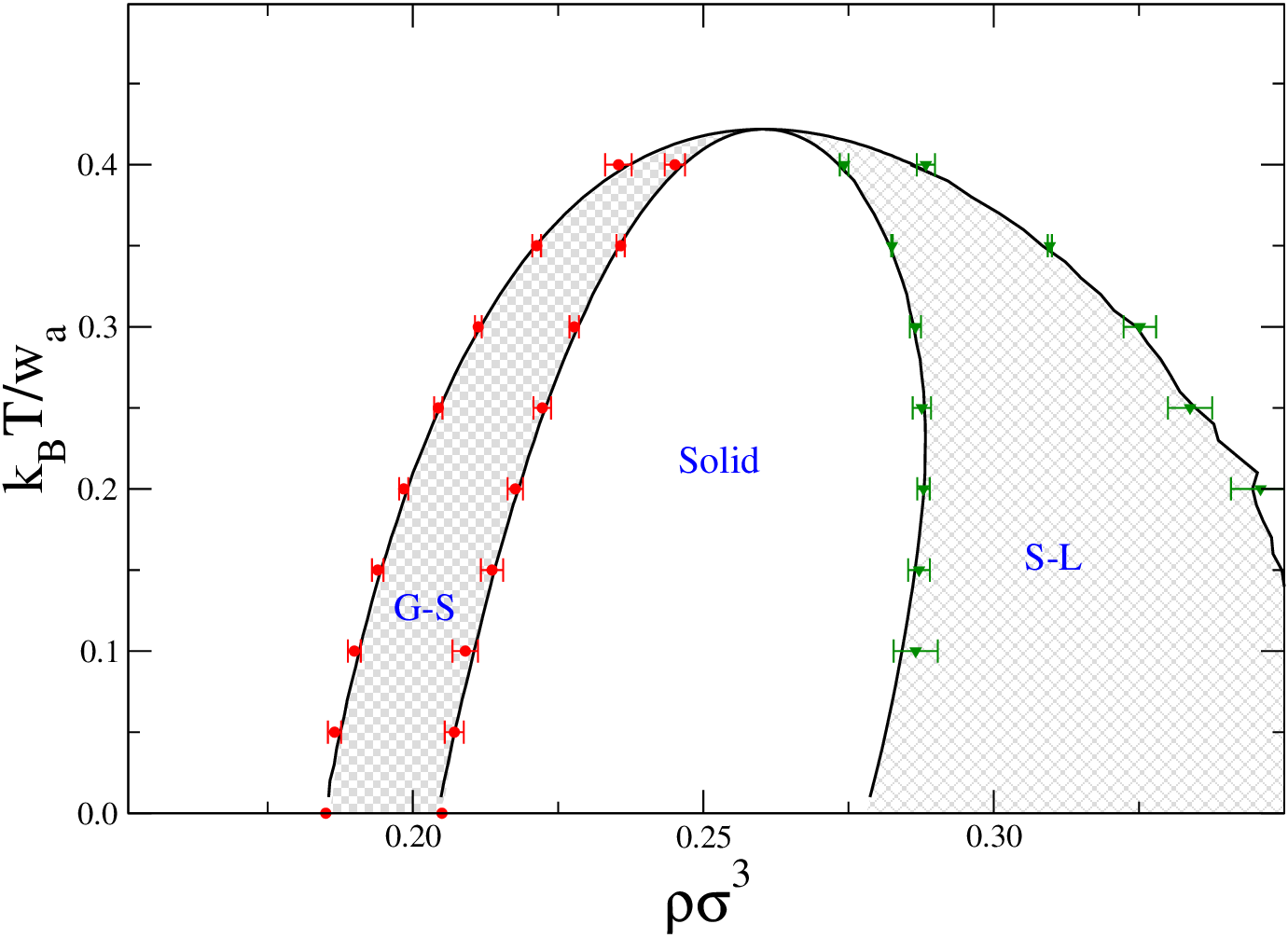}
\caption{(Colour online)  $\rho$-T phase diagram of the purely repulsive 2SRP model as
  obtained from computer simulation. Points: thermodynamic integration, 
lines: Gibbs-Duhem integration. Legend as in Fig.\protect\ref{fig8}.}
\label{fig10}
\end{figure}

When the attractive interactions are switched off, even if we have seen
that features such as the curves of compressibility and heat capacity maxima are
hardly affected, the phase diagram simplifies substantially. Obviously, the LV
transition disappears, and one is left with a rather peculiar fluid-solid
transition (see Fig.~\ref{fig10}). First, one observes that when density is
increased along an isotherm starting from the low density fluid, the systems
crystallises into an {\it fcc} phase, which melts upon further compression.
At higher density values one could find the rhombohedric, cubic and
hexagonal phases predicted by Jagla\cite{JCP_1999_111_8980}. This melting upon
compression is similar to the behaviour of water. In the 2SRP model
this is a purely energy driven transition: for certain densities the interparticle
distance is necessarily $r < d_a$ and the repulsive spheres overlap, thus
an ordered configuration no longer  corresponds to an energy minimum.
At sufficiently large densities ($\rho\sigma^3>0.9$), the hard cores
will regain  their
controlling role and an entropy driven crystallisation will take place (into either {\it fcc}
or {\it hcp} structures). In the intermediate region ($\rho\sigma^3>0.35$) the
 interplay between entropy and energy will give rise to a much more complex
 scenario with different solid phases in equilibrium with the fluid phase.

 Finally, from the location of the fluid-solid equilibrium curve, it is now clear that a
 possible LL transition would be preempted by crystallisation. When comparing
 Figures \ref{fig8} and \ref{fig10} one can see  that the effect of the
 dispersive interactions is to increase the stability of  the liquid (fluid) phase with
 respect to the solid. Otherwise, the LL critical temperature happens
 to be fairly close to the the maximum temperature at which the
 fluid-solid equilibrium takes place ($T_{LL}^*=0.38$ vs $T_{FS}^*=0.42$).

\section{Conclusions}
In summary, we have presented a detailed study of both the attractive and the
purely repulsive two scale ramp potential models, using both a self consistent
integral equation (HMSA) and different Monte Carlo techniques. From this study,
it becomes clear that the use of a thermodynamically consistent closure is
essential if one wishes to capture the presence of the LL transition in the attractive model
and the density anomaly of the repulsive fluid. Moreover, this type of model
clearly highlights the shortcomings of local consistency criteria, such as those
usually implemented in the HMSA. Unfortunately, our calculations using a global
consistency condition on the pressure calculated from the virial and the
isothermal compressibility only converge  at high temperatures, well away from
the LL and density anomaly region. This implies, that the closure is missing
some essential features of the physical behaviour in the region controlled by
the soft repulsive interactions. In order to bypass this limitation one should
explore the use of other functional forms\cite{JCP_2000_112_810} or use an
integro-differential approach of the Self Consistent Ornstein Zernike
Approximation (SCOZA) \cite{JCP_1977_67_439,MP_1984_52_1071} which has been
recently implemented for systems with bounded
potentials\cite{JCP_2006_124_064503}. Despite these limitations, the
theoretical approach yields quantitatively correct estimates of the LV
transition, the density anomaly and predicts the existence of a LL transition
(although with a substantial underestimation of the LL critical density). 

By means of extensive Monte Carlo simulations, we have unveiled the rather complicated
shape  of a significant part of the fluid-solid diagram for the A2SRP
models, in which three triple points have been detected. In the purely
repulsive model, one finds that the crystallisation preempts the liquid-liquid
phase separation. This system has, in
common with water, a solid phase that melts upon compression.
In the intermediate region of the phase diagram, in which the hard
core the the repulsive ramp are competing (i.e. when neither entropy nor energy
but a subtle combination of both  magnitudes leads the system behaviour) one
should expect even more complex phase diagrams for both attractive and
repulsive models. This will be the subject of future work.

\acknowledgments
The authors acknowledge support from the Direcci\'on General de
Investigaci\'on Cient\'\i fica y T\'ecnica under grant
no. FIS2004-02954-C03-01 and the Direcci\'on General de
Universidades e Investigaci\'on de la Comunidad de Madrid under Grant
S0505/ESP/0299, program MOSSNOHO-CM.
One of the authors (C. McBride) would like to thank the  CSIC for the award of an I3P post-doctoral contract,
co-financed by the European Social Fund.


\begin{thebibliography}{55}
\expandafter\ifx\csname natexlab\endcsname\relax\def\natexlab#1{#1}\fi
\expandafter\ifx\csname bibnamefont\endcsname\relax
  \def\bibnamefont#1{#1}\fi
\expandafter\ifx\csname bibfnamefont\endcsname\relax
  \def\bibfnamefont#1{#1}\fi
\expandafter\ifx\csname citenamefont\endcsname\relax
  \def\citenamefont#1{#1}\fi
\expandafter\ifx\csname url\endcsname\relax
  \def\url#1{\texttt{#1}}\fi
\expandafter\ifx\csname urlprefix\endcsname\relax\def\urlprefix{URL }\fi
\providecommand{\bibinfo}[2]{#2}
\providecommand{\eprint}[2][]{\url{#2}}

\bibitem[{\citenamefont{Tanaka}(1996)}]{JCP_1996_105_5099}
\bibinfo{author}{\bibfnamefont{H.}~\bibnamefont{Tanaka}}, \bibinfo{journal}{J.
  Chem. Phys.} \textbf{\bibinfo{volume}{105}}, \bibinfo{pages}{5099}
  (\bibinfo{year}{1996}).

\bibitem[{\citenamefont{Yamada et~al.}(2002)\citenamefont{Yamada, Mossa,
  Stanley, and Sciortino}}]{PRL_2002_88_195701}
\bibinfo{author}{\bibfnamefont{M.}~\bibnamefont{Yamada}},
  \bibinfo{author}{\bibfnamefont{S.}~\bibnamefont{Mossa}},
  \bibinfo{author}{\bibfnamefont{H.~E.} \bibnamefont{Stanley}},
  \bibnamefont{and}
  \bibinfo{author}{\bibfnamefont{F.}~\bibnamefont{Sciortino}},
  \bibinfo{journal}{Phys. Rev. Lett.} \textbf{\bibinfo{volume}{88}},
  \bibinfo{pages}{195701} (\bibinfo{year}{2002}).

\bibitem[{\citenamefont{Xu et~al.}(2005)\citenamefont{Xu, Kumar, Buldyrev,
  Chen, Poole, Sciortino, and Stanley}}]{PNAS_2005_102_16558}
\bibinfo{author}{\bibfnamefont{L.}~\bibnamefont{Xu}},
  \bibinfo{author}{\bibfnamefont{P.}~\bibnamefont{Kumar}},
  \bibinfo{author}{\bibfnamefont{S.~V.} \bibnamefont{Buldyrev}},
  \bibinfo{author}{\bibfnamefont{S.}~\bibnamefont{Chen}},
  \bibinfo{author}{\bibfnamefont{P.~H.} \bibnamefont{Poole}},
  \bibinfo{author}{\bibfnamefont{F.}~\bibnamefont{Sciortino}},
  \bibnamefont{and} \bibinfo{author}{\bibfnamefont{H.~E.}
  \bibnamefont{Stanley}}, \bibinfo{journal}{Proc. Natl. Acad. Sci. U.S.A.}
  \textbf{\bibinfo{volume}{102}}, \bibinfo{pages}{16558}
  (\bibinfo{year}{2005}).

\bibitem[{\citenamefont{Brovchenko et~al.}(2005)\citenamefont{Brovchenko,
  Geiger, and Oleinikova}}]{JCP_2005_123_044515}
\bibinfo{author}{\bibfnamefont{I.}~\bibnamefont{Brovchenko}},
  \bibinfo{author}{\bibfnamefont{A.}~\bibnamefont{Geiger}}, \bibnamefont{and}
  \bibinfo{author}{\bibfnamefont{A.}~\bibnamefont{Oleinikova}},
  \bibinfo{journal}{J. Chem. Phys.} \textbf{\bibinfo{volume}{123}},
  \bibinfo{pages}{044515} (\bibinfo{year}{2005}).

\bibitem[{\citenamefont{Vega and Abascal}(2005)}]{JCP_2005_123_144504}
\bibinfo{author}{\bibfnamefont{C.}~\bibnamefont{Vega}} \bibnamefont{and}
  \bibinfo{author}{\bibfnamefont{J.~L.~F.} \bibnamefont{Abascal}},
  \bibinfo{journal}{J. Chem. Phys.} \textbf{\bibinfo{volume}{123}},
  \bibinfo{pages}{144504} (\bibinfo{year}{2005}).

\bibitem[{\citenamefont{Katayama et~al.}(2000)\citenamefont{Katayama, Mizutani,
  Utsumi, Shimomura, Yamakata, and K.-ichi Funakoshi}}]{NAT_2000_403_170}
\bibinfo{author}{\bibfnamefont{Y.}~\bibnamefont{Katayama}},
  \bibinfo{author}{\bibfnamefont{T.}~\bibnamefont{Mizutani}},
  \bibinfo{author}{\bibfnamefont{W.}~\bibnamefont{Utsumi}},
  \bibinfo{author}{\bibfnamefont{O.}~\bibnamefont{Shimomura}},
  \bibinfo{author}{\bibfnamefont{M.}~\bibnamefont{Yamakata}}, \bibnamefont{and}
  \bibinfo{author}{\bibfnamefont{K.}~\bibnamefont{-ichi Funakoshi}},
  \bibinfo{journal}{Nature (London)} \textbf{\bibinfo{volume}{403}},
  \bibinfo{pages}{170} (\bibinfo{year}{2000}).

\bibitem[{\citenamefont{Tanaka et~al.}(2004)\citenamefont{Tanaka, Kurita, and
  Mataki}}]{PRL_2004_92_025701}
\bibinfo{author}{\bibfnamefont{H.}~\bibnamefont{Tanaka}},
  \bibinfo{author}{\bibfnamefont{R.}~\bibnamefont{Kurita}}, \bibnamefont{and}
  \bibinfo{author}{\bibfnamefont{H.}~\bibnamefont{Mataki}},
  \bibinfo{journal}{Phys. Rev. Lett.} \textbf{\bibinfo{volume}{92}},
  \bibinfo{pages}{025701} (\bibinfo{year}{2004}).

\bibitem[{\citenamefont{Kurita and Tanaka}(2005)}]{JPCM_2005_17_L293}
\bibinfo{author}{\bibfnamefont{R.}~\bibnamefont{Kurita}} \bibnamefont{and}
  \bibinfo{author}{\bibfnamefont{H.}~\bibnamefont{Tanaka}},
  \bibinfo{journal}{J.Phys.: Condens. Matter} \textbf{\bibinfo{volume}{17}},
  \bibinfo{pages}{L293} (\bibinfo{year}{2005}).

\bibitem[{\citenamefont{Sastry and Angell}(2003)}]{NM_2003_2_739}
\bibinfo{author}{\bibfnamefont{S.}~\bibnamefont{Sastry}} \bibnamefont{and}
  \bibinfo{author}{\bibfnamefont{C.~A.} \bibnamefont{Angell}},
  \bibinfo{journal}{Nat. Matter.} \textbf{\bibinfo{volume}{2}},
  \bibinfo{pages}{739} (\bibinfo{year}{2003}).

\bibitem[{\citenamefont{Ghiringhelli et~al.}(2004)\citenamefont{Ghiringhelli,
  Los, Meijer, Fasolino, and Frenkel}}]{PRB_2004_69_100101}
\bibinfo{author}{\bibfnamefont{L.~M.} \bibnamefont{Ghiringhelli}},
  \bibinfo{author}{\bibfnamefont{J.~H.} \bibnamefont{Los}},
  \bibinfo{author}{\bibfnamefont{E.~J.} \bibnamefont{Meijer}},
  \bibinfo{author}{\bibfnamefont{A.}~\bibnamefont{Fasolino}}, \bibnamefont{and}
  \bibinfo{author}{\bibfnamefont{D.}~\bibnamefont{Frenkel}},
  \bibinfo{journal}{Phys. Rev. B} \textbf{\bibinfo{volume}{69}},
  \bibinfo{pages}{100101} (\bibinfo{year}{2004}).

\bibitem[{\citenamefont{Saika-Voivod et~al.}(2000)\citenamefont{Saika-Voivod,
  Sciortino, , and Poole}}]{PRE_2000_63_011202}
\bibinfo{author}{\bibfnamefont{I.}~\bibnamefont{Saika-Voivod}},
  \bibinfo{author}{\bibfnamefont{F.}~\bibnamefont{Sciortino}}, ,
  \bibnamefont{and} \bibinfo{author}{\bibfnamefont{P.~H.} \bibnamefont{Poole}},
  \bibinfo{journal}{Phys. Rev. E} \textbf{\bibinfo{volume}{63}},
  \bibinfo{pages}{011202} (\bibinfo{year}{2000}).

\bibitem[{\citenamefont{Mishima}(1993)}]{JCP_1993_100_5910}
\bibinfo{author}{\bibfnamefont{O.}~\bibnamefont{Mishima}}, \bibinfo{journal}{J.
  Chem. Phys.} \textbf{\bibinfo{volume}{100}}, \bibinfo{pages}{5910}
  (\bibinfo{year}{1993}).

\bibitem[{\citenamefont{Mukherjee et~al.}(2001)\citenamefont{Mukherjee, Vaidya,
  and Sugandhi}}]{PRL_2001_87_195501}
\bibinfo{author}{\bibfnamefont{G.~D.} \bibnamefont{Mukherjee}},
  \bibinfo{author}{\bibfnamefont{S.~N.} \bibnamefont{Vaidya}},
  \bibnamefont{and} \bibinfo{author}{\bibfnamefont{V.}~\bibnamefont{Sugandhi}},
  \bibinfo{journal}{Phys. Rev. Lett.} \textbf{\bibinfo{volume}{87}},
  \bibinfo{pages}{195501} (\bibinfo{year}{2001}).

\bibitem[{\citenamefont{Smith et~al.}(1995)\citenamefont{Smith, Shero,
  Chizmeshya, and Wolf}}]{JCP_1995_102_6851}
\bibinfo{author}{\bibfnamefont{K.~H.} \bibnamefont{Smith}},
  \bibinfo{author}{\bibfnamefont{E.}~\bibnamefont{Shero}},
  \bibinfo{author}{\bibfnamefont{A.}~\bibnamefont{Chizmeshya}},
  \bibnamefont{and} \bibinfo{author}{\bibfnamefont{G.~H.} \bibnamefont{Wolf}},
  \bibinfo{journal}{J. Chem. Phys.} \textbf{\bibinfo{volume}{102}},
  \bibinfo{pages}{6851} (\bibinfo{year}{1995}).

\bibitem[{\citenamefont{Cohen et~al.}(1996)\citenamefont{Cohen, Ha, Zhao, Lee,
  Fischer, Strouse, and Kivelson}}]{JPC_1996_100_8518}
\bibinfo{author}{\bibfnamefont{I.}~\bibnamefont{Cohen}},
  \bibinfo{author}{\bibfnamefont{A.}~\bibnamefont{Ha}},
  \bibinfo{author}{\bibfnamefont{X.}~\bibnamefont{Zhao}},
  \bibinfo{author}{\bibfnamefont{M.}~\bibnamefont{Lee}},
  \bibinfo{author}{\bibfnamefont{T.}~\bibnamefont{Fischer}},
  \bibinfo{author}{\bibfnamefont{M.~J.} \bibnamefont{Strouse}},
  \bibnamefont{and} \bibinfo{author}{\bibfnamefont{D.}~\bibnamefont{Kivelson}},
  \bibinfo{journal}{J. Phys. Chem.} \textbf{\bibinfo{volume}{100}},
  \bibinfo{pages}{8518} (\bibinfo{year}{1996}).

\bibitem[{\citenamefont{Roberts and Debenedetti}(1996)}]{JCP_1996_105_658}
\bibinfo{author}{\bibfnamefont{C.~J.} \bibnamefont{Roberts}} \bibnamefont{and}
  \bibinfo{author}{\bibfnamefont{P.~G.} \bibnamefont{Debenedetti}},
  \bibinfo{journal}{J. Chem. Phys.} \textbf{\bibinfo{volume}{105}},
  \bibinfo{pages}{658} (\bibinfo{year}{1996}).

\bibitem[{\citenamefont{Roberts et~al.}(1996)\citenamefont{Roberts,
  Panagiotopoulos, and Debenedetti}}]{PRL_1996_77_4386}
\bibinfo{author}{\bibfnamefont{C.~J.} \bibnamefont{Roberts}},
  \bibinfo{author}{\bibfnamefont{A.~Z.} \bibnamefont{Panagiotopoulos}},
  \bibnamefont{and} \bibinfo{author}{\bibfnamefont{P.~G.}
  \bibnamefont{Debenedetti}}, \bibinfo{journal}{Phys. Rev. Lett.}
  \textbf{\bibinfo{volume}{77}}, \bibinfo{pages}{4386 } (\bibinfo{year}{1996}).

\bibitem[{\citenamefont{Hemmer and Stell}(1970)}]{PRL_1970_24_1284}
\bibinfo{author}{\bibfnamefont{P.~C.} \bibnamefont{Hemmer}} \bibnamefont{and}
  \bibinfo{author}{\bibfnamefont{G.}~\bibnamefont{Stell}},
  \bibinfo{journal}{Phys. Rev. Lett.} \textbf{\bibinfo{volume}{24}},
  \bibinfo{pages}{1284 } (\bibinfo{year}{1970}).

\bibitem[{\citenamefont{Stell and Hemmer}(1972)}]{JCP_1972_56_4274}
\bibinfo{author}{\bibfnamefont{G.}~\bibnamefont{Stell}} \bibnamefont{and}
  \bibinfo{author}{\bibfnamefont{P.}~\bibnamefont{Hemmer}},
  \bibinfo{journal}{J. Chem. Phys.} \textbf{\bibinfo{volume}{56}},
  \bibinfo{pages}{4274} (\bibinfo{year}{1972}).

\bibitem[{\citenamefont{Skibinsky et~al.}(2004)\citenamefont{Skibinsky,
  Buldyrev, Franzese, Malescio, and Stanley}}]{PRE_2004_69_061206}
\bibinfo{author}{\bibfnamefont{A.}~\bibnamefont{Skibinsky}},
  \bibinfo{author}{\bibfnamefont{S.~V.} \bibnamefont{Buldyrev}},
  \bibinfo{author}{\bibfnamefont{G.}~\bibnamefont{Franzese}},
  \bibinfo{author}{\bibfnamefont{G.}~\bibnamefont{Malescio}}, \bibnamefont{and}
  \bibinfo{author}{\bibfnamefont{H.~E.} \bibnamefont{Stanley}},
  \bibinfo{journal}{Phys. Rev. E} \textbf{\bibinfo{volume}{69}},
  \bibinfo{pages}{061206} (\bibinfo{year}{2004}).

\bibitem[{\citenamefont{Tarjus et~al.}(2003)\citenamefont{Tarjus,
  CAlba-Simionesco, Grousson, Viot, and Kivelson}}]{JPCM_2003_15_S1077}
\bibinfo{author}{\bibfnamefont{G.}~\bibnamefont{Tarjus}},
  \bibinfo{author}{\bibnamefont{CAlba-Simionesco}},
  \bibinfo{author}{\bibfnamefont{M.}~\bibnamefont{Grousson}},
  \bibinfo{author}{\bibfnamefont{P.}~\bibnamefont{Viot}}, \bibnamefont{and}
  \bibinfo{author}{\bibfnamefont{D.}~\bibnamefont{Kivelson}},
  \bibinfo{journal}{J.Phys.: Condens. Matter} \textbf{\bibinfo{volume}{15}},
  \bibinfo{pages}{S1077} (\bibinfo{year}{2003}).

\bibitem[{\citenamefont{Jagla}(1999)}]{JCP_1999_111_8980}
\bibinfo{author}{\bibfnamefont{E.~A.} \bibnamefont{Jagla}},
  \bibinfo{journal}{J. Chem. Phys.} \textbf{\bibinfo{volume}{111}},
  \bibinfo{pages}{8980} (\bibinfo{year}{1999}).

\bibitem[{\citenamefont{Xu et~al.}(2006)\citenamefont{Xu, Ehrenberg, Buldyrev,
  and Stanley}}]{JPCM_2006_18_S2239}
\bibinfo{author}{\bibfnamefont{L.}~\bibnamefont{Xu}},
  \bibinfo{author}{\bibfnamefont{I.}~\bibnamefont{Ehrenberg}},
  \bibinfo{author}{\bibfnamefont{S.~V.} \bibnamefont{Buldyrev}},
  \bibnamefont{and} \bibinfo{author}{\bibfnamefont{H.~E.}
  \bibnamefont{Stanley}}, \bibinfo{journal}{J. Phys.: Condens. Matter}
  \textbf{\bibinfo{volume}{18}}, \bibinfo{pages}{S2239} (\bibinfo{year}{2006}).

\bibitem[{\citenamefont{Gibson and Wilding}(2006)}]{PRE_2006_73_061507}
\bibinfo{author}{\bibfnamefont{H.~M.} \bibnamefont{Gibson}} \bibnamefont{and}
  \bibinfo{author}{\bibfnamefont{N.~B.} \bibnamefont{Wilding}},
  \bibinfo{journal}{Phys. Rev. E} \textbf{\bibinfo{volume}{73}},
  \bibinfo{pages}{061507} (\bibinfo{year}{2006}).

\bibitem[{\citenamefont{Caballero and Puertas}(2006)}]{PRE_2006_74_051506}
\bibinfo{author}{\bibfnamefont{J.~B.} \bibnamefont{Caballero}}
  \bibnamefont{and} \bibinfo{author}{\bibfnamefont{A.~M.}
  \bibnamefont{Puertas}}, \bibinfo{journal}{Phys. Rev. E}
  \textbf{\bibinfo{volume}{74}}, \bibinfo{pages}{051506}
  (\bibinfo{year}{2006}).

\bibitem[{\citenamefont{Kumar et~al.}(2005)\citenamefont{Kumar, Buldyrev,
  Sciortino, Zaccarelli, and Stanley}}]{PRE_2005_72_021501}
\bibinfo{author}{\bibfnamefont{P.}~\bibnamefont{Kumar}},
  \bibinfo{author}{\bibfnamefont{S.~V.} \bibnamefont{Buldyrev}},
  \bibinfo{author}{\bibfnamefont{F.}~\bibnamefont{Sciortino}},
  \bibinfo{author}{\bibfnamefont{E.}~\bibnamefont{Zaccarelli}},
  \bibnamefont{and} \bibinfo{author}{\bibfnamefont{H.~E.}
  \bibnamefont{Stanley}}, \bibinfo{journal}{Phys. Rev. E}
  \textbf{\bibinfo{volume}{72}}, \bibinfo{pages}{021501}
  (\bibinfo{year}{2005}).

\bibitem[{\citenamefont{Sharma et~al.}(2006)\citenamefont{Sharma, Chakraborty,
  and Chakravarty}}]{JCP_2006_125_204501}
\bibinfo{author}{\bibfnamefont{R.}~\bibnamefont{Sharma}},
  \bibinfo{author}{\bibfnamefont{S.~N.} \bibnamefont{Chakraborty}},
  \bibnamefont{and}
  \bibinfo{author}{\bibfnamefont{C.}~\bibnamefont{Chakravarty}},
  \bibinfo{journal}{J. Chem. Phys.} \textbf{\bibinfo{volume}{125}},
  \bibinfo{pages}{204501} (\bibinfo{year}{2006}).

\bibitem[{\citenamefont{Errington et~al.}(2006)\citenamefont{Errington,
  Truskett, and Mittal}}]{JCP_2006_125_244502}
\bibinfo{author}{\bibfnamefont{J.~R.} \bibnamefont{Errington}},
  \bibinfo{author}{\bibfnamefont{T.~M.} \bibnamefont{Truskett}},
  \bibnamefont{and} \bibinfo{author}{\bibfnamefont{J.}~\bibnamefont{Mittal}},
  \bibinfo{journal}{J. Chem. Phys.} \textbf{\bibinfo{volume}{125}},
  \bibinfo{pages}{244502} (\bibinfo{year}{2006}).

\bibitem[{\citenamefont{Netz et~al.}(2006)\citenamefont{Netz, Buldyrev,
  Barbosa, and Stanley}}]{PRE_2006_73_061504}
\bibinfo{author}{\bibfnamefont{P.~A.} \bibnamefont{Netz}},
  \bibinfo{author}{\bibfnamefont{S.~V.} \bibnamefont{Buldyrev}},
  \bibinfo{author}{\bibfnamefont{M.~C.} \bibnamefont{Barbosa}},
  \bibnamefont{and} \bibinfo{author}{\bibfnamefont{H.~E.}
  \bibnamefont{Stanley}}, \bibinfo{journal}{Phys. Rev. E}
  \textbf{\bibinfo{volume}{73}}, \bibinfo{pages}{061504}
  (\bibinfo{year}{2006}).

\bibitem[{\citenamefont{Yan et~al.}(2005)\citenamefont{Yan, Buldyrev,
  Giovambattista, and Stanley}}]{PRL_2005_95_130604}
\bibinfo{author}{\bibfnamefont{Z.}~\bibnamefont{Yan}},
  \bibinfo{author}{\bibfnamefont{S.~V.} \bibnamefont{Buldyrev}},
  \bibinfo{author}{\bibfnamefont{N.}~\bibnamefont{Giovambattista}},
  \bibnamefont{and} \bibinfo{author}{\bibfnamefont{H.~E.}
  \bibnamefont{Stanley}}, \bibinfo{journal}{Phys. Rev. Lett.}
  \textbf{\bibinfo{volume}{95}}, \bibinfo{pages}{130604}
  (\bibinfo{year}{2005}).

\bibitem[{\citenamefont{Errington and Debenedetti}(2001)}]{NAT_2001_409_318}
\bibinfo{author}{\bibfnamefont{J.~R.} \bibnamefont{Errington}}
  \bibnamefont{and} \bibinfo{author}{\bibfnamefont{P.~G.}
  \bibnamefont{Debenedetti}}, \bibinfo{journal}{Nature (London)}
  \textbf{\bibinfo{volume}{409}}, \bibinfo{pages}{318} (\bibinfo{year}{2001}).

\bibitem[{\citenamefont{Errington et~al.}(2003)\citenamefont{Errington,
  Debenedetti, and Torquato}}]{JCP_2003_118_2256}
\bibinfo{author}{\bibfnamefont{J.~R.} \bibnamefont{Errington}},
  \bibinfo{author}{\bibfnamefont{P.~G.} \bibnamefont{Debenedetti}},
  \bibnamefont{and} \bibinfo{author}{\bibfnamefont{S.}~\bibnamefont{Torquato}},
  \bibinfo{journal}{J. Chem. Phys.} \textbf{\bibinfo{volume}{118}},
  \bibinfo{pages}{2256} (\bibinfo{year}{2003}).

\bibitem[{\citenamefont{Rosenfeld}(1977)}]{PRA_1977_15_002545}
\bibinfo{author}{\bibfnamefont{Y.}~\bibnamefont{Rosenfeld}},
  \bibinfo{journal}{Phys. Rev. A} \textbf{\bibinfo{volume}{15}},
  \bibinfo{pages}{2545} (\bibinfo{year}{1977}).

\bibitem[{\citenamefont{Rosenfeld}(1999)}]{JPCM_1999_11_05415}
\bibinfo{author}{\bibfnamefont{Y.}~\bibnamefont{Rosenfeld}},
  \bibinfo{journal}{J. Phys.: Condens. Matter} \textbf{\bibinfo{volume}{11}},
  \bibinfo{pages}{5415} (\bibinfo{year}{1999}).

\bibitem[{\citenamefont{Dzugutov}(1996)}]{N_1996_381_00137_photocopy}
\bibinfo{author}{\bibfnamefont{M.}~\bibnamefont{Dzugutov}},
  \bibinfo{journal}{Nature (London)} \textbf{\bibinfo{volume}{381}},
  \bibinfo{pages}{137} (\bibinfo{year}{1996}).

\bibitem[{\citenamefont{Weeks et~al.}(1971)\citenamefont{Weeks, Chandler, and
  Andersen}}]{JCP_1971_54_5237}
\bibinfo{author}{\bibfnamefont{J.~D.} \bibnamefont{Weeks}},
  \bibinfo{author}{\bibfnamefont{D.}~\bibnamefont{Chandler}}, \bibnamefont{and}
  \bibinfo{author}{\bibfnamefont{H.~C.} \bibnamefont{Andersen}},
  \bibinfo{journal}{J. Chem. Phys.} \textbf{\bibinfo{volume}{54}},
  \bibinfo{pages}{5237} (\bibinfo{year}{1971}).

\bibitem[{\citenamefont{Ornstein and Zernike}(1914)}]{KNAW_1914_17_0793}
\bibinfo{author}{\bibfnamefont{L.~S.} \bibnamefont{Ornstein}} \bibnamefont{and}
  \bibinfo{author}{\bibfnamefont{F.}~\bibnamefont{Zernike}},
  \textbf{\bibinfo{volume}{17}}, \bibinfo{pages}{793} (\bibinfo{year}{1914}).

\bibitem[{\citenamefont{Morita and Hiroik}(23)}]{PTP_1960_023_1003}
\bibinfo{author}{\bibfnamefont{T.}~\bibnamefont{Morita}} \bibnamefont{and}
  \bibinfo{author}{\bibfnamefont{K.}~\bibnamefont{Hiroik}},
  \textbf{\bibinfo{volume}{1003}}, \bibinfo{pages}{1960} (\bibinfo{year}{23}).

\bibitem[{\citenamefont{Franzese et~al.}(2002)\citenamefont{Franzese, Malescio,
  Skibinsky, Buldyrev, and Stanley}}]{PRE_2002_66_051206}
\bibinfo{author}{\bibfnamefont{G.}~\bibnamefont{Franzese}},
  \bibinfo{author}{\bibfnamefont{G.}~\bibnamefont{Malescio}},
  \bibinfo{author}{\bibfnamefont{A.}~\bibnamefont{Skibinsky}},
  \bibinfo{author}{\bibfnamefont{S.~V.} \bibnamefont{Buldyrev}},
  \bibnamefont{and} \bibinfo{author}{\bibfnamefont{H.~E.}
  \bibnamefont{Stanley}}, \bibinfo{journal}{Phys. Rev. E}
  \textbf{\bibinfo{volume}{66}}, \bibinfo{pages}{051206}
  (\bibinfo{year}{2002}).

\bibitem[{\citenamefont{Martynov et~al.}(1999)\citenamefont{Martynov, Sarkisov,
  and Vompe}}]{JCP_1999_110_3961}
\bibinfo{author}{\bibfnamefont{G.~A.} \bibnamefont{Martynov}},
  \bibinfo{author}{\bibfnamefont{G.~N.} \bibnamefont{Sarkisov}},
  \bibnamefont{and} \bibinfo{author}{\bibfnamefont{A.~G.} \bibnamefont{Vompe}},
  \bibinfo{journal}{J. Chem. Phys.} \textbf{\bibinfo{volume}{110}},
  \bibinfo{pages}{3961} (\bibinfo{year}{1999}).

\bibitem[{\citenamefont{Hansen and Zerah}(1985)}]{PL_1985_108A_277}
\bibinfo{author}{\bibfnamefont{J.-P.} \bibnamefont{Hansen}} \bibnamefont{and}
  \bibinfo{author}{\bibfnamefont{G.}~\bibnamefont{Zerah}},
  \bibinfo{journal}{Phys. Lett. A} \textbf{\bibinfo{volume}{108A}},
  \bibinfo{pages}{277} (\bibinfo{year}{1985}).

\bibitem[{\citenamefont{Zerah and Hansen}(1986)}]{JCP_1986_84_2336}
\bibinfo{author}{\bibfnamefont{G.}~\bibnamefont{Zerah}} \bibnamefont{and}
  \bibinfo{author}{\bibfnamefont{J.-P.} \bibnamefont{Hansen}},
  \bibinfo{journal}{J. Chem. Phys.} \textbf{\bibinfo{volume}{84}},
  \bibinfo{pages}{2336} (\bibinfo{year}{1986}).

\bibitem[{\citenamefont{Rogers and Young}(1984)}]{PRA_1984_30_999}
\bibinfo{author}{\bibfnamefont{F.~J.} \bibnamefont{Rogers}} \bibnamefont{and}
  \bibinfo{author}{\bibfnamefont{D.~A.} \bibnamefont{Young}},
  \bibinfo{journal}{Phys. Rev. A} \textbf{\bibinfo{volume}{30}},
  \bibinfo{pages}{999 } (\bibinfo{year}{1984}).

\bibitem[{\citenamefont{Caccamo and Pellicane}(2002)}]{JCP_2002_117_5072}
\bibinfo{author}{\bibfnamefont{C.}~\bibnamefont{Caccamo}} \bibnamefont{and}
  \bibinfo{author}{\bibfnamefont{G.}~\bibnamefont{Pellicane}},
  \bibinfo{journal}{J. Chem. Phys.} \textbf{\bibinfo{volume}{117}},
  \bibinfo{pages}{5072} (\bibinfo{year}{2002}).

\bibitem[{\citenamefont{Caccamo et~al.}(1993)\citenamefont{Caccamo, Giaquinta,
  and Giunta}}]{JPCM_1993_5_B75}
\bibinfo{author}{\bibfnamefont{C.}~\bibnamefont{Caccamo}},
  \bibinfo{author}{\bibfnamefont{P.}~\bibnamefont{Giaquinta}},
  \bibnamefont{and} \bibinfo{author}{\bibfnamefont{G.}~\bibnamefont{Giunta}},
  \bibinfo{journal}{J. Phys.: Condens. Matter} \textbf{\bibinfo{volume}{5}},
  \bibinfo{pages}{B75} (\bibinfo{year}{1993}).
\bibitem[{\citenamefont{Belloni}(1988)}]{JCP_1988_88_5143}
\bibinfo{author}{\bibfnamefont{L.}~\bibnamefont{Belloni}}, \bibinfo{journal}{J.
  Chem. Phys.} \textbf{\bibinfo{volume}{88}}, \bibinfo{pages}{5143}
  (\bibinfo{year}{1988}).

\bibitem[{\citenamefont{Sarkisov and Lomba}(2005)}]{JCP_2005_122_214504}
\bibinfo{author}{\bibfnamefont{G.}~\bibnamefont{Sarkisov}} \bibnamefont{and}
  \bibinfo{author}{\bibfnamefont{E.}~\bibnamefont{Lomba}}, \bibinfo{journal}{J.
  Chem. Phys.} \textbf{\bibinfo{volume}{122}}, \bibinfo{pages}{214504}
  (\bibinfo{year}{2005}).


\bibitem{PRE_2005_71_046132}
E. Lomba, C. Mart{\'\i}n, N. G. Almarza and F. Lado, Phys Rev. E {\bf 71}, 046132 (2005).

\bibitem{PRL_2001_86_2050}
Fugao Wang and D. P. Landau, Phys. Rev. Lett. {\bf 86}, 2050 (2001),

\bibitem{PRE_2001_64_056101}
Fugao Wang and D. P. Landau, Phys Rev. E {\bf 64}, 56101 (2001).

\bibitem{frenkel_smit}
D. Frenkel and B. Smit, Understanding Molecular Simulation, From Algorithms to
Application, (Academic Press, London, 2002)


\bibitem{mp_1993_78_1331}
D. A. Kofke, Mol. Phys. {\bf 78}, 1331 (1993)

\bibitem{jcp_1993_98_4149}
D. A. Kofke, J. Chem. Phys. {\bf 98}, 4149 (1993)

\bibitem{allen_tildesley}
M.P. Allen and D.J. Tidesley, Computer Simulation of Liquids (Clarendon Press, Oxford, 1987) 

\bibitem{jpcm_1997_9_8951} 
R. J. Speedy, J. Phys.: Condensed Matter {\bf 9}, 8951 (1997)

\bibitem{jcp_1969_51_635} 
N. F. Carnahan and K. E. Starling, J. Chem. Phys. {\bf 51} , 635 (1969)

\bibitem{jcp_1998_10_4387}
R. J. Speedy, J. Phys.: Condensed Matter {\bf 10}, 4387 (1998)


\bibitem[{\citenamefont{Sarkisov}(2003)}]{JCP_2003_119_373}
\bibinfo{author}{\bibfnamefont{G.~N.} \bibnamefont{Sarkisov}},
  \bibinfo{journal}{J. Chem. Phys.} \textbf{\bibinfo{volume}{119}},
  \bibinfo{pages}{373} (\bibinfo{year}{2003}).

\bibitem[{\citenamefont{Lomba}(1989)}]{MP_1989_68_87}
\bibinfo{author}{\bibfnamefont{E.}~\bibnamefont{Lomba}}, \bibinfo{journal}{Mol.
  Phys.} \textbf{\bibinfo{volume}{68}}, \bibinfo{pages}{87 }
  (\bibinfo{year}{1989}).

\bibitem[{\citenamefont{Sanz et~al.}(2004)\citenamefont{Sanz, Vega, Abascal,
  and MacDowell}}]{PRL_2004_92_255701}
\bibinfo{author}{\bibfnamefont{E.}~\bibnamefont{Sanz}},
  \bibinfo{author}{\bibfnamefont{C.}~\bibnamefont{Vega}},
  \bibinfo{author}{\bibfnamefont{J.~L.~F.} \bibnamefont{Abascal}},
  \bibnamefont{and} \bibinfo{author}{\bibfnamefont{L.~G.}
  \bibnamefont{MacDowell}}, \bibinfo{journal}{Phys. Rev. Lett.}
  \textbf{\bibinfo{volume}{92}}, \bibinfo{pages}{255701}
  (\bibinfo{year}{2004}).

\bibitem[{\citenamefont{Sastry et~al.}(1996)\citenamefont{Sastry, Debenedetti,
  Sciortino, and Stanley}}]{PRE_1996_53_6144}
\bibinfo{author}{\bibfnamefont{S.}~\bibnamefont{Sastry}},
  \bibinfo{author}{\bibfnamefont{P.~G.} \bibnamefont{Debenedetti}},
  \bibinfo{author}{\bibfnamefont{F.}~\bibnamefont{Sciortino}},
  \bibnamefont{and} \bibinfo{author}{\bibfnamefont{H.~E.}
  \bibnamefont{Stanley}}, \bibinfo{journal}{Phys. Rev. E}
  \textbf{\bibinfo{volume}{53}}, \bibinfo{pages}{6144 } (\bibinfo{year}{1996}).

\bibitem[{\citenamefont{Fernaud et~al.}(2000)\citenamefont{Fernaud, Lomba, and
  Lee}}]{JCP_2000_112_810}
\bibinfo{author}{\bibfnamefont{M.-J.} \bibnamefont{Fernaud}},
  \bibinfo{author}{\bibfnamefont{E.}~\bibnamefont{Lomba}}, \bibnamefont{and}
  \bibinfo{author}{\bibfnamefont{L.~L.} \bibnamefont{Lee}},
  \bibinfo{journal}{J. Chem. Phys.} \textbf{\bibinfo{volume}{112}},
  \bibinfo{pages}{810} (\bibinfo{year}{2000}).

\bibitem[{\citenamefont{H{\o}ye and Stell}(1977)}]{JCP_1977_67_439}
\bibinfo{author}{\bibfnamefont{J.~S.} \bibnamefont{H{\o}ye}} \bibnamefont{and}
  \bibinfo{author}{\bibfnamefont{G.}~\bibnamefont{Stell}}, \bibinfo{journal}{J.
  Chem. Phys.} \textbf{\bibinfo{volume}{67}}, \bibinfo{pages}{439}
  (\bibinfo{year}{1977}).

\bibitem[{\citenamefont{H{\o}ye and Stell}(1984)}]{MP_1984_52_1071}
\bibinfo{author}{\bibfnamefont{J.}~\bibnamefont{H{\o}ye}} \bibnamefont{and}
  \bibinfo{author}{\bibfnamefont{G.}~\bibnamefont{Stell}},
  \bibinfo{journal}{Mol.Phys.} \textbf{\bibinfo{volume}{52}},
  \bibinfo{pages}{1071 } (\bibinfo{year}{1984}).

\bibitem[{\citenamefont{Mladek et~al.}(2006)\citenamefont{Mladek, Kahl, and
  Neumann}}]{JCP_2006_124_064503}
\bibinfo{author}{\bibfnamefont{B.~M.} \bibnamefont{Mladek}},
  \bibinfo{author}{\bibfnamefont{G.}~\bibnamefont{Kahl}}, \bibnamefont{and}
  \bibinfo{author}{\bibfnamefont{M.}~\bibnamefont{Neumann}},
  \bibinfo{journal}{J. Chem. Phys.} \textbf{\bibinfo{volume}{124}},
  \bibinfo{pages}{064503} (\bibinfo{year}{2006}).

\end{thebibliography}

\end{document}